\documentclass[preprint2]{aastex}
\usepackage{epsfig}
\usepackage{amsfonts,amssymb,amsmath}

\makeatletter
\newcommand\footnoteref[1]{\protected@xdef\@thefnmark{\ref{#1}}\@footnotemark}
\makeatother

\def\lsim{\mathrel{\rlap{\lower3.5pt\hbox{\hskip0.5pt$\sim$}}
    \raise0.5pt\hbox{$<$}}}                
\def\gsim{~\rlap{$>$}{\lower 1.0ex\hbox{$\sim$}}}
\shortauthors{Romeo et al.} \shorttitle{RS scatter}

\def\ltsim{\mathrel{\rlap{\lower 3pt\hbox{$\sim$}}\raise
2.0pt\hbox{$<$}}}
\def\gtsim{\mathrel{\rlap{\lower 3pt\hbox{$\sim$}} \raise
2.0pt\hbox{$>$}}}

\newcommand{\q}{\begin{equation}}
\newcommand{\qa}{\begin{eqnarray}}
\newcommand{\qs}{\begin{eqnarray*}}
\newcommand{\nq}{\end{equation}}
\newcommand{\nqa}{\end{eqnarray}}
\newcommand{\nqs}{\end{eqnarray*}}
\def\be{\begin{equation}}
\def\ee{\end{equation}}

\def\lsim{\mathrel{\rlap{\lower3.5pt\hbox{\hskip0.5pt$\sim$}}
    \raise0.5pt\hbox{$<$}}}                
\def\gsim{~\rlap{$>$}{\lower 1.0ex\hbox{$\sim$}}}

\begin{document}
\title{The high-redshift evolution of the Red Sequence scatter from joint simulations and HAWK-I Cluster Survey.}

\author{A.D. Romeo \altaffilmark{1}, P. Cerulo\altaffilmark{2,3}, Kang Xi\altaffilmark{1}, E. Contini \altaffilmark{1}, J. Sommer-Larsen\altaffilmark{4,5}, I. Gavignaud\altaffilmark{6}}

\altaffiltext{1}{
Purple Mountain Observatory, 2 Beijing Xilu, 210008 Nanjing, China}\email{\texttt
romeo@pmo.ac.cn, pcerulo@astro-udec.cl, kangxi@pmo.ac.cn}
\altaffiltext{2}{Universidad de Concepci\'on, Chile}
\altaffiltext{3}{Centre for Astrophysics and Supercomputing, Swinburne University of Technology, PO Box 218, Hawthorn, VIC 3122, Australia}
\altaffiltext{4}{Excellence Cluster Universe, Technische Universit$\ddot{a}$t M$\ddot{u}$nchen, Boltzmannstr. 2, D-85748 Garching bei M$\ddot{a}$nchen, Germany}
\altaffiltext{5}{Dark Cosmology Centre, Niels Bohr Institute, University of Copenhagen, Juliane Maries Vej 30, DK-2100 Copenhagen, Denmark}
\altaffiltext{6}{Departamento de Ciencias Fisicas, Universidad Andres Bello, Avda. Republica 252, Santiago, Chile}

\begin{abstract}
We study the evolution of the Red Sequence (RS) scatter in galaxy clusters and groups simultaneously using predictions from our
simulations (cosmological hydrodynamic + semi-analytical) as well as observational data from the HAWK-I Cluster Survey (HCS),
a sample of galaxy clusters at redshifts $0.8 < z < 1.5$.
We analyze the intrinsic scatter of the RS to investigate whether the stellar age can be its main effective driver, 
at the same time assessing the role of metallicity variations in shaping the RS building at around epoch $z\sim$1 and beyond.
To this purpose we rely on various methods to derive the average age and age spread from the RS colour scatter of the HCS sample, with the aid of population synthesis models.
The RS scatter predicted by the models at $z\lsim 0.7$ is found to not depend on the star formation history adopted, whilst
at $z\gsim 0.7$ more gradually decaying star-formation laws result in larger RS scatters.
In general the correlation found between age and rest-frame colour scatters is quite robust, although all age scatter estimations ultimately 
depend on the definition of RS as well as on the completeness limits adopted.
We find that the age spread of RS galaxies predicted by both hydrodynamical simulations and SAM increases with cosmic epoch, while the ratio between the age spread and the average age remains approximately constant. Both trends are in agreement with observational results from both the HCS and other literature samples.
\end{abstract}

\keywords{galaxies: evolution --- galaxies: clusters: general --- galaxies: groups: general --- galaxies: star formation}

\section{Introduction}

The slope and scatter of the colour-magnitude (CM) relation provide relevant constraints upon the epoch and duration of the star formation in early-type (ET) galaxies. 
The theoretical interpretation of the CM relation along the Red Sequence (RS) which best explains the observational results is that this is a mass--metallicity sequence, with the most massive and red galaxies being also the most metal rich (see Kodama \& Arimoto 1997, Gladders et al. 1998). 
A variation in time of the RS slope will then be attributable to a differential chemical enrichment along the sequence which preferentially
yields more metals in the faint galaxies, resulting into a steeper slope at higher redshift (see Gallazzi et al. 2006).
Yet, many works from the recent literature report that there has been little evolution in the slope of the RS in clusters at redshifts $0<z<1.8$ 
(e.g.:\ Ellis et al. 1997, Lidman et al. 2008, Mei et al. 2009, Andreon 2014). 

An intrinsic evolution of the RS slope might also be due to
bluer, younger galaxies that have been quenched during their infall to the cluster core and are thus entering the RS
after recent SF episodes (e.g. De Lucia et al. 2007): these objects will redden more
rapidly than the previous older and brighter RS members, leading to a flattening
of the RS at progressively lower redshifts. 

While the CMR slope mirrors a metallicity sequence, 
the colour scatter of the galaxies along the RS is always the combination of the scatter in both their ages and metallicities,
what makes of it a highly degenerate observable.
Its main origin is commonly thought to be the spread in galaxy ages at given luminosity or mass, that is the heterogeneity of star formation histories and durations between galaxies (see Bower, Lucey \& Ellis 1992, hereafter BLE92).
Yet, even though the colour scatter primarily constrains the variance in the age differences among galaxies at a given epoch
(see Kodama \& Arimoto 1997), still it cannot be ruled out a metallicity contribution (Nelan et al. 2005).

Indeed there is evidence that also the variation of the slope may be sensitive to age differences between more and less
massive galaxies (see Kodama et al. 1998), even though such a trend (at the base of the ``archaeological" downsizing)
is not clear and seems to depend on the presence of an external heating source like AGN.
Romeo et al. (2008) showed that the epoch at which star-forming galaxies turn to passive corresponds to
the epoch at which the RS slope becomes null, that is also when its scatter becomes that of the pure quiescent population (so called ``Dead Sequence").

The intrinsic scatter of the CMR in the local Universe is found to remain quite tight (again BLE92) 
and this implies that either galaxies in nearby clusters had formed at high $z$, or through quite synchronized formation epochs.
Several morphologically-selected samples have been studied (e.g. Arag\'on Salamanca et al. 1993),
mostly pointing at a constancy of the scatter with redshift, and a mean colour reddening
consistent with passive ageing of old stellar populations (e.g. Ellis et al. 1997).
A tightness in scatter at higher redshift was highlighted by e.g. Ellis et al. (1997) at $z$=0.55 and Stanford,
Eisenhardt \& Dickinson (1998) at $z$=0.9, implying a homogeneity of the ET population across the cosmic time which may
contrast with the strong evolution of the cluster galaxies predicted by the Butcher-Oemler effect (1984).
The latter is interpreted by means of the transformation of blue field galaxies into red cluster ones, but only if this infall is
accompanied by a morphological change from spirals to ETs (or an equivalent change from star-forming to passive), 
then the scatter of the CMR is expected to increase, as a result of the higher number of blue ETs approaching the RS.

If the ET fraction were not to evolve with redshift, one would expect the scatter of the RS to increase
with lookback time, just because mirroring the increasing relative age spread of the galaxy sample.
But dynamical interactions and mergers are likely to break such a one-to-one correspondence between
the samples at higher $z$ and at present: in the same cluster (or group),
the non-star forming galaxies at high $z$ are a subsample of the same objects at $z$=0, giving rise to the
so-called ``progenitor bias" when comparing ET galaxies belonging to clusters at different redshifts.
In van Dokkum et al. (1998), the scatter of the RS is discussed as a combined constraint on the
spread in galaxy ages and the mean age of the galaxies at a given $z$, so that can be expressed as
proportional to the ratio $\sigma_\tau/<\tau>$. Both quantities should increase
with time, in particular the former because of the broadening of the RS itself. Hence, the
scatter in colour of the RS evolves with redshift depending on the difference in the rates at which 
both $\sigma_\tau$ and $<\tau>$ decrease with $z$. Stanford et al. (1998) reported a nearly constant
scatter in the CMR up to $z\simeq$1, which would imply that the mean galaxy age increases in time at
the same rate as the scatter in ages during that period.

Most of the semi-analytical (SAM) and hydrodynamical simulations have been unable to reproduce both the size of the intrinsic 
scatter and its lack of evolution with redshift.
In early SAMs the scatter decreased with redshift (by a factor of twice up to $z$=1.5 in the rest-frame U-V) because the
selection of galaxies at high redshift was biased towards very early epochs of galaxy formation (Kauffmann \& Charlot 1998).
In hydro-dynamical simulations (Romeo et al. 2008), a similar decrease in
the scatter was found for galaxies belonging to the {\it Dead Sequence}, where they stopped forming stars.
In the more recent SAM of Menci et al. (2008), the average scatter (up to $z$=1.5) is almost independent of redshift, 
but still remains twice to thrice larger than observed, partly due to the RS selection and pointing at a possible observational bias towards
most evolved clusters. 

In this paper we will also explore the possibility that models are missing some physical process leading to the small scatter observed, by addressing the following broadlines:
\begin{itemize}
\item {Derivation of stellar age scatter from the observed RS colour scatter following different techniques;}
\item {comparison of observed age scatter with predictions from simulations;}
\item {comparison of the observed age scatter with predictions from stellar population evolution models;}
\item {comparison of predicted age scatter in simulations as a function of stellar metallicity;}
\item {comparison of the sigma/tau estimated from the observations with that predicted in simulations.}
\end{itemize}

\section{Methods}

The formation of galaxies in our simulations (hereafter SIM) is followed ab initio within a 150 Mpc cosmological volume, which allows us to account for their interaction with a large-scale environment. 
This simulation serves as input for the volume re-normalization technique to achieve higher resolution in zoomed
regions such as individual clusters and groups, which are later re-simulated with the full hydrodynamical code (see Romeo et al. 2008). 
By re-zooming on target selected regions at higher resolution, we resolved down to galaxy-sized haloes, which allows to model their stellar populations:
a mass resolution of stellar and gas particles down to to $3\times 10^7 h^{-1}M_\odot$ was achieved in this way.
In particular, the virial mass of the haloes selected at $z=0$ to be re-simulated at higher resolution are $1\times 10^{14} M_\odot$ for groups,
$3\times 10^{14} M_\odot$ for the lower mass cluster and $1.2\times 10^{15} M_\odot$ for the larger one - measured within the virial radius defined as enclosing an over-density of 200 times the critical background density. Out  of the 12 re-simulated groups at z=0, four were of fossil nature (see D'Onghia et al. 2005).
Galaxies from homogeneous density regions were stacked together to make up four environmental classes: cluster cores within $1/3 R_{200}$ (IN), cluster outskirts outside $1/3 R_{200}$ up to $R_{200}$ (OUT), normal groups (NG), and fossil groups (FG) -the latters undivided.

The SAM used in this paper is based on Kang et al. (2005), which later was developed in Kang et al. (2012), and to which we refer for details. 
The cosmological box had a size of 200$Mpc/h$ on each side and was populated with $1024^{3}$ particles.
From the initial simulation box, a total number of haloes above the threshold of $1\times 10^{14}M_{\odot}$ were extracted, ranging from 2830 ($z$=0) to 2150 ($z$=0.5), 1300 ($z$=1), down to 300 ($z$=2); out of these, those below $1.5\times 10^{14}M_{\odot}$ were classified as groups.
The merger trees are constructed by following the subhaloes resolved in FOF haloes at each snapshot. The SAM is then grafted on the merger trees and self-consistently models the physics processes governing galaxy formation, such as gas cooling, star formation, supernova, and AGN feedback. Finally, the galaxy luminosity and colours are calculated based on the stellar population synthesis of Bruzual \& Charlot (2003) adopting a Chabrier stellar initial mass function (IMF) (Chabrier 2003). 

In particular, the SAM includes radio-mode AGN feedback, parametrized in a phenomenological way. 
This mechanism works so as to suppress the
cooling in massive haloes, whose star formation in the central galaxies is hence shut off, drifting them onto the RS.
In the present model, developed in Kang, Jing \& Silk (2006), the gas cooling continues to form stars until a massive spheroid forms at the galaxy centre. 
In this way, more massive and luminous galaxies
can be formed at high redshift, thus achieving a successful reproduction of the observed LF in rest-frame {\it K }band and the galaxy colour distributions.

On the low-mass end, the treatment of gas stripping from satellites in the model is not instantaneous (see Kang \& van den Bosch 2008), what helps to
alleviate the problem of the overabundance of red faint galaxies, which has generally been ascribed to the low efficiency in tidal disruption of dwarfs and at the same time to a too efficient ram-pressure stripping of those dwarfs that survived. In our model the outer hot gas is let be gradually stripped over a timescale of about 3 Gyr: this prolonged strangulation results in a longer-lasting cooling and hence delayed quenching of SF. 
Moreover, the current  version of the SAM presented in this paper has also implemented the formation of diffuse inter-galactic light (IGL) from stripping channel possibly subsequent to merger events. 

The cosmology adopted thoroughout is $\Lambda CDM$, with $\Omega_\Lambda$=0.7/0.73, $\Omega_m$=0.30/0.27$, \Omega_{b}$=0.045, $h$=0.7, and $\sigma_8$=0.9/0.81 in the SIM and SAM respectively.
All magnitudes are quoted in the Vega system if not stated otherwise in the text.

\begin{table}[t!]
  \caption{The HAWK-I Cluster Survey (HCS) sample with the clusters listed in order of increasing redshift. The DM halo masses $M_{DM}$ were estimated by Jee et al. (2011) through a weak lensing analysis on HST ACS images. The last column gives the $V_{AB}$-band magnitude limit (see text).}

  \begin{tabular}{|l|c|c|c|}
    \hline
     \multicolumn{1}{|c}{Cluster} & \multicolumn{1}{|c}{Redshift} & \multicolumn{1}{|c|}{$M_{DM}$} & \multicolumn{1}{c|}{$V_{AB}$} \\
     \multicolumn{1}{|c}{} & \multicolumn{1}{|c}{} & \multicolumn{1}{|c|}{($10^{14} M_\odot$)} & \multicolumn{1}{c|}{(mag)} \\
    \hline
     \hline
     RX0152 & 0.84 & $4.4^{+0.7}_{{-}{0.5}}$  & -18.0 \\
     RCS2319 & 0.91 & $5.8^{+2.3}_{{-}1.6}$ & -17.4 \\
     XMM1229  & 0.98 & $5.3^{+1.7}_{{-}1.2}$ & -19.4 \\
     RCS0220 & 1.03 & $4.8^{+1.8}_{{-}1.3}$ & -18.3 \\
     RCS2345 & 1.04 & $2.4^{+1.1}_{{-}0.7}$ & -18.4 \\
     XMMU0223 & 1.22 & $7.4^{+2.5}_{{-}1.8}$ & -20.2 \\
     RDCS1252 & 1.24 & $6.8^{+1.2}_{{-}1.0}$ & -18.3 \\
     XMMU2235 & 1.39 & $7.3^{+1.7}_{{-}1.4}$ & -19.2 \\
     XMMXCS2215 & 1.45 & $4.3^{+3.0}_{{-}1.7}$ & -20.7 \\
  \hline
  \end{tabular}
\label{table1_PC}
\end{table}

\section{Data and Observations}\label{sec:data_and_observations_PC}

We use  a sample of galaxy clusters at redshifts $0.8 < z < 1.5$ drawn from the HAWK-I Cluster Survey (HCS, Lidman et al. 2013). The sample is extensively discussed in Cerulo et al. (2016, hereafter C2016), while this section summarises its main properties. 
The HCS is a programme aimed at studying galaxy clusters in the redshift range $0.8<z<1.6$, which represents the epoch at which most of the star formation is thought to have been suppressed in cluster cores (see e.g. Hilton et al. 2010, Tran et al. 2010, Webb et al. 2013, Brodwin et al. 2013, Alberts et al. 2014, Santos et al. 2014, Fassbender et al. 2014, Santos et al. 2015, although see Andreon et al. 2014 for a different perspective).

Clusters were observed with the near infrared (NIR) High Acuity Wide-field K-band Imager (HAWK-I, Pirard et al. 2004) at the ESO Very Large Telescope (VLT) in Chile in the J and Ks bands. Images cover a field of $\sim 10' \times 10'$ (corresponding to a projected distance of 4.8 Mpc at $z = 1.0$), reaching an image quality, parametrised by the full width at half maximum (FWHM) of the point spread function (PSF), of $FWHM = 0.3''-0.4''$ and image depth, estimated by the 90\% magnitude completeness limit, in the range 22.5-25.0 AB mag in both bands.
C2016 selected 9 galaxy clusters in the HCS sample for which Hubble Space Telescope (HST) imaging data taken with the Advanced Camera for Surveys (ACS) were available. Additional archival optical and NIR data were used, together with spectra from various observing programs, to study the evolution of the RS within a projected distance of $0.5 \times R_{200}$ from each cluster centre. The sample is summarised in Table \ref{table1_PC}.

C2016 analyzed the RS in these clusters, estimating the intercept (zero-point), slope, and intrinsic scatter of the CM relation, by
using all the spectroscopically confirmed cluster members and adopting a statistical background subtraction technique to account for field contamination, 
which becomes significant at faint magnitudes (i.e.: ACS F850LP $\sim$ 23.0-25.0 AB mag). 

In summary this method, which closely follows Pimbblet et al. (2002) and Valentinuzzi et al. (2011), consists in estimating the ratio of the number of galaxies observed in the cluster field within a range of colour and magnitude (a {\itshape{colour-magnitude cell}}) to the number of galaxies observed in the same colour-magnitude cell in a comparison field. Such a ratio defines the probability for a galaxy in that cell to be in the field ($P_{field}$). 
C2016 ran 200 Monte Carlo simulations in each of which a random number $0.0 < P < 1.0$ was assigned to each galaxy on the RS. This number was compared with $P_{field}$ and all galaxies with $P \leq P_{field}$ were rejected from the sample as field interlopers. In this way, in each iteration a {\itshape{clean}} sample is created and the CM relation is fitted to it.

The linear fit to the RS is then performed with a robust linear fitting technique employing the Tukey's biweight function. The uncertainties on the zero-point ($a$) and slope ($b$) are estimated as the $1 \sigma$ width of the bootstrap distributions of these parameters after 1000 iterations. The intrinsic scatter $\sigma_c$ is determined, following Lidman et al. (2008) and Mei et al. (2009), as the residual scatter that needs to be added to the photometric error in order to obtain reduced $\chi^2 = 1.0$. The uncertainty on $\sigma_c$ is estimated with the same procedure used for zero-point and slope.

The zero-point, slope, and intrinsic scatter measured on the RS in the observer frame were converted to their rest-frame equivalent in the $B-V$ vs $V$, $U-V$ vs $V$, and $U-B$ vs $B$ colour-magnitude planes. The conversion procedure is based on the method outlined in Appendix B of Mei et al. (2009) and consists in generating a sample of models extracted from the Bruzual \& Charlot (2003, hereafter BC2003) library with three values of metallicity (0.4, 1, 2.5 $Z_\odot$), formation redshifts in the range $2.0 < z_f < 5.0$, two star-formation laws (single instantaneous burst and exponentially declining with e-folding time $\tau = 1.0$ Gyr), and Salpeter (1955) IMF. We derived the colours and magnitudes predicted by these models in both the observer frame and in the rest frame at the redshift of each cluster. Linear models were then fitted to the relationships between observer- and rest-frame quantities and were used in the conversions to absolute magnitudes and colours. The rest-frame zero-points, slopes, and intrinsic scatters for the HCS clusters in the $B-V$ vs $V$, $U-V$ vs $V$, and $U-B$ vs $B$ are all tabulated in C2016.

The RS parameters were estimated using all the galaxies down to the 90\% magnitude completeness limit in each cluster. Although this approach allowed the exploitation of the maximum information achievable in each cluster, it also resulted in a heterogeneous sampling of the RS, which is highlighted in Table 1 (last column). Furthermore, as discussed above, the intrinsic RS scatter is estimated as the residual scatter to be added to the photometric error in order to obtain reduced $\chi^2 = 1$, meaning that larger photometric errors may systematically reduce the value of the intrinsic scatter.

In order to evaluate this effect, we repeated the fit to the RS using a fixed magnitude limit for all the clusters in the HCS. This limit corresponds to the V-band absolute magnitude at the 90\% completeness limit in the field of the cluster XMM J2215-1738 (XMMXCS2215, $z = 1.46$), which is the shallowest data set in the HCS. This selection resulted in a stellar mass limited sample which we hereafter name the {\itshape{common limit}} sample. The implications of the different selections on the estimate of the intrinsic color scatter of the RS will be discussed in Section 4.1.

\begin{figure*}	
\centering
\includegraphics[width=8cm,clip]{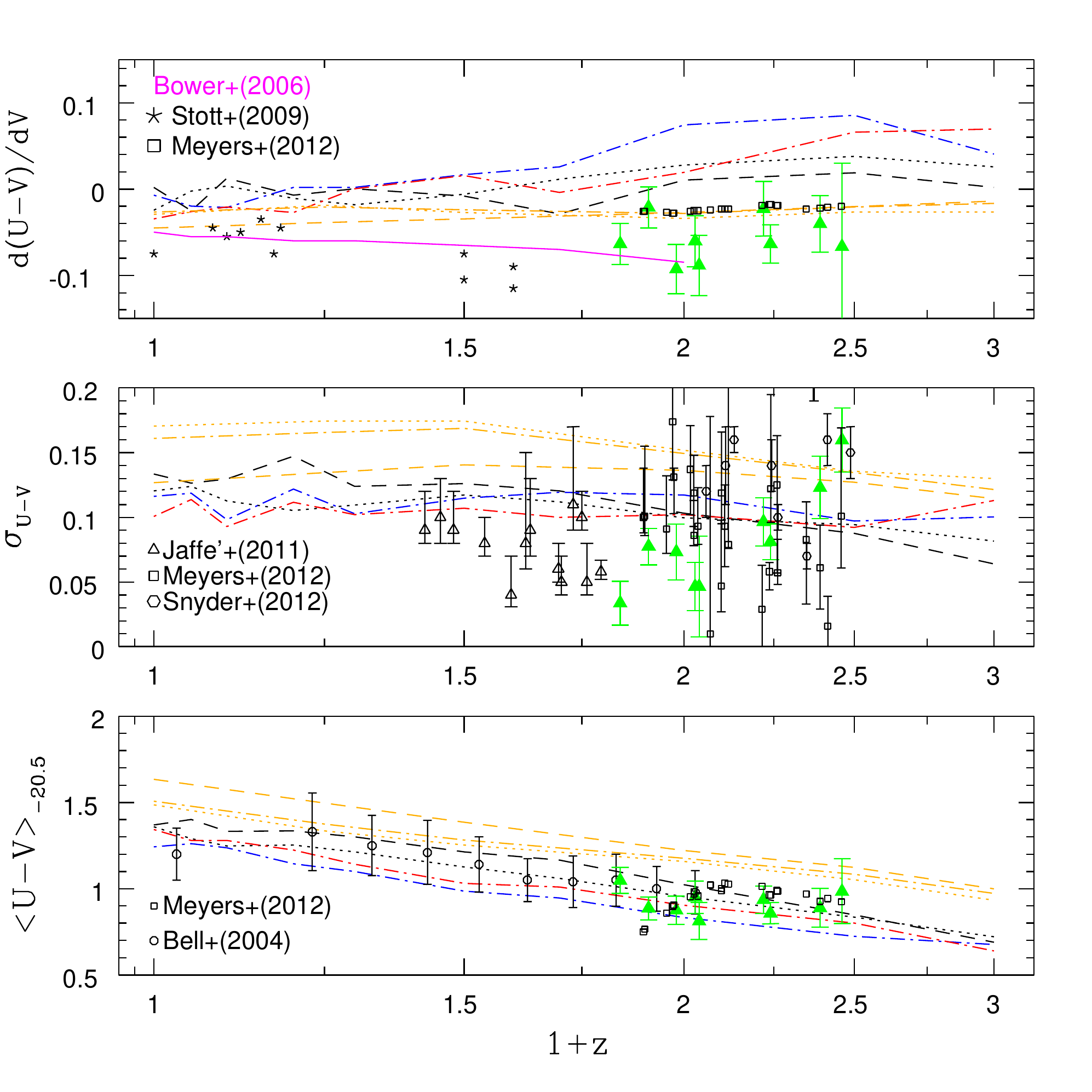}
\hfill 
\includegraphics[width=8cm,clip]{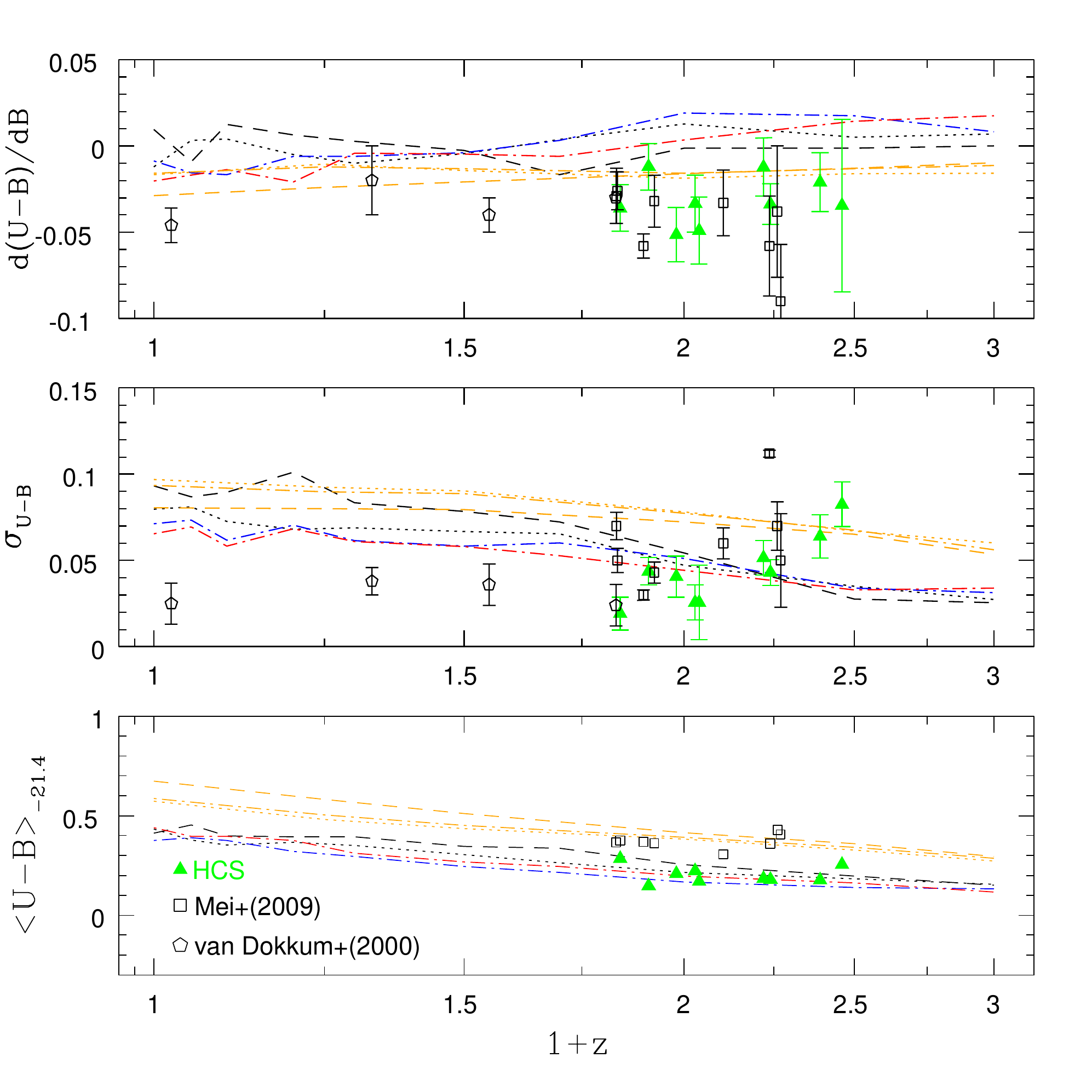}
\caption{Slope, scatter and zero-point (measured at $M_V$=-20.5 or $M_B$=-21.4 mag) from the linear fit to CMR, in different rest-frame bands, of the 
RS samples of 
cluster core (dashed), cluster outskirts (dotted), normal groups (blue dot-dashed) and fossil groups (red dot-dashed) galaxies from SIM, 
compared with the analogous regions from SAM (orange lines) and our HCS data (green triangles: solid 90\% completeness limit, empty common completeness limit). 
More observational data for clusters are also shown, as well as another semi-analytical prediction for the U-V slope (see text). 
Data from Jaff\'e et al. (2011) refer to observed frame V-I or R-I, roughly correspondent to U-V at each cluster's redshift.
}
\includegraphics[width=8cm,clip]{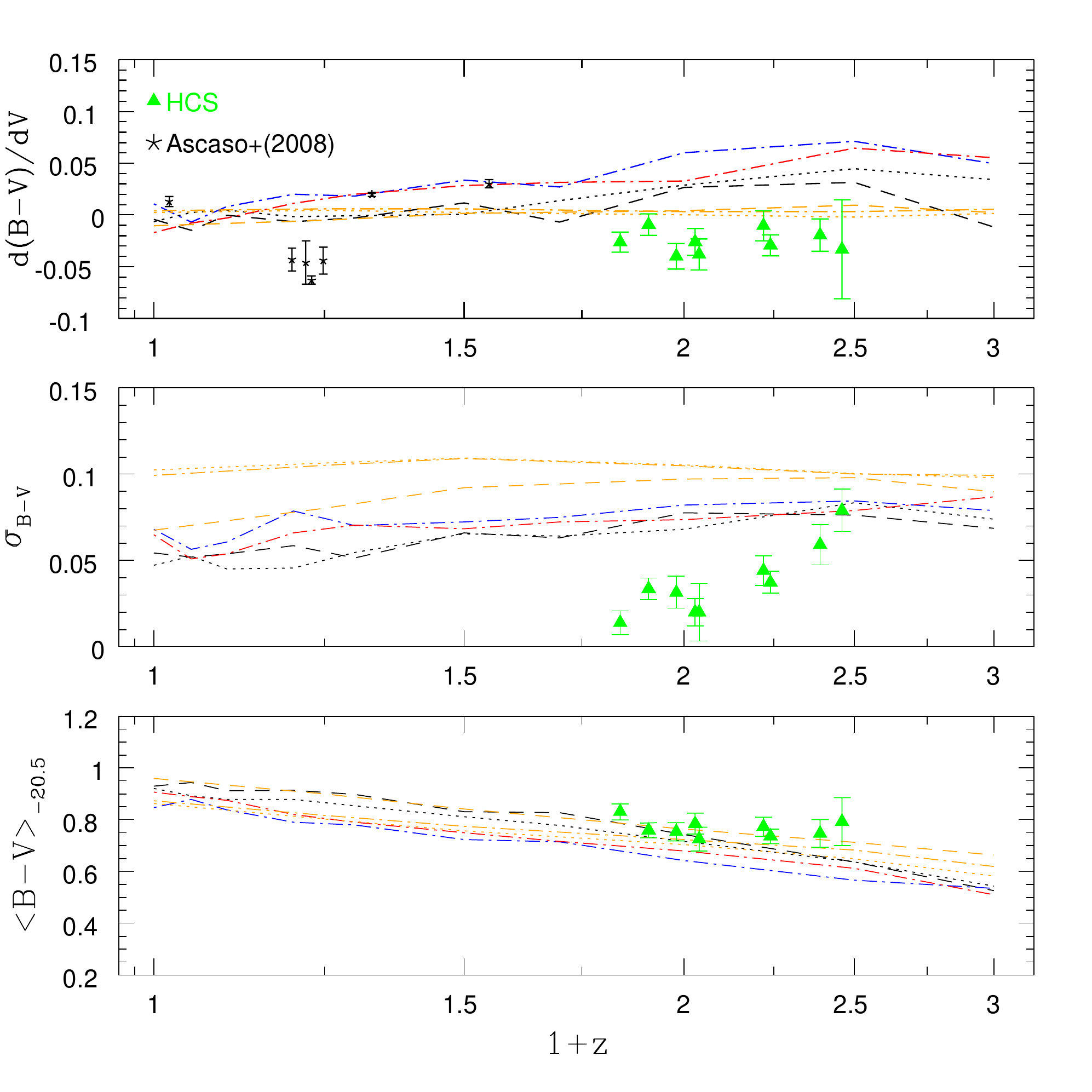}
\hfill
\includegraphics[width=8cm,clip]{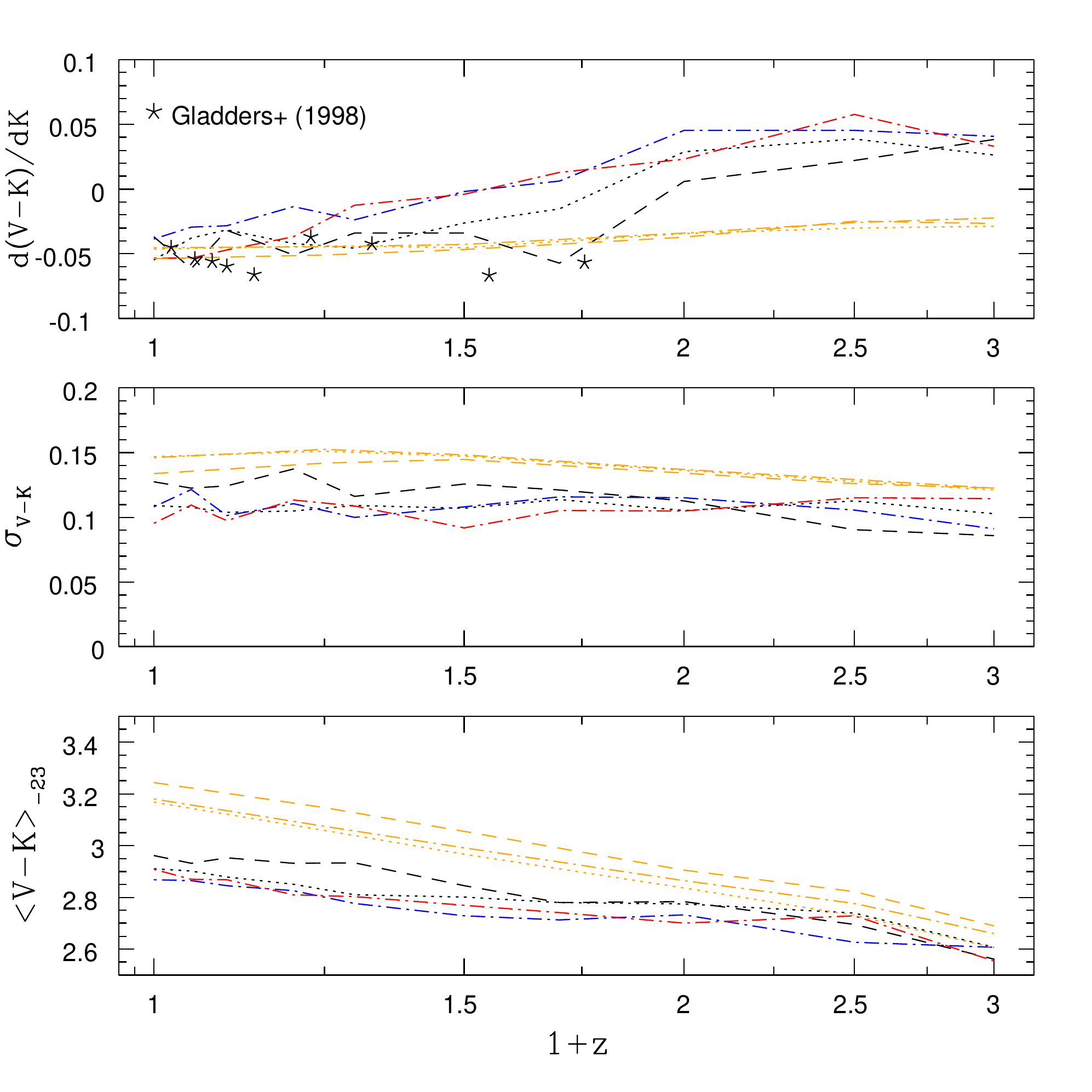}
\label{slosca}
\end{figure*}

\section{Results}

\subsection{RS parameters}

In Fig. \ref{slosca} we show the slope, scatter and intercept (zero-point) of the linear bestfit to the RS samples from both SIM and SAM in different rest-frame colours, along with those from the HCS clusters. A caveat is that HCS data refer to half of $R_{200}$ aperture, so that should be compared with both "IN" and "OUT" curves from models, which are separated by a threshold of one-third of $R_{200}$.
As general feature, both slope and scatter are higher when involving {\it U} band, because the rest-frame ({\it U-V}) straddles the 4000 A break, thus providing a very age-sensitive broadband colour that depends on the possible presence of young stellar populations, contributing to increase the scatter.
Regarding the BCGs, we tested that their inclusion in the models significatively affects the slope evolution, while the scatter does not depend on them, as expected. In the following we will not include BCGs in either SIM or SAM, while in the HCS this is included as not outstanding from the global RS distribution. 

Our previous dataset of simulated galaxies (Romeo et al. 2008) predicted a stronger evolution of the slope of the RS, becoming positive at $z\gsim1$, which was not in agreement with many observational results. Adopting the same selection criteria introduced in Romeo et al. (2015), the slope fairly lowers yet remains still high at $z\gsim1.2$ and in all cases is steeper than in the SAM, in every colour. 
In fact, all the SAM galaxies follow a quasi-constant slope independent of the environment, while in SIM a larger spread is noticed between groups and clusters.
With respect to other data, our HCS as well as SAM keep a fairly flat trend of the slope, in agreement with Mei et al. (2009) in ({\it U-B}), but not with Stott et al. (2009) in ({\it U-V}), whose slope gets more negative, that is increasingly steeper, with $z$; and are marginally consistent with the SAM reconstructed by the latter authors by adopting the prescription in Bower et al. (2006), which also yields a marked, although less steep than their own data, slope evolution.

In general, SAM presents overall lower values of the slope but higher values of the scatter ($\simeq$0.1-0.15 in {\it U-V}) with respect to SIM, in all colours and classes. Moreover, all the SAM zero-points are redder than the SIM ones and also respect to all observational points. 
This leads to the result that our HCS data (green triangles) are best reproduced by the SAM when referring to the slope, and by the SIM when scatter and zero-point are concerned. Since our SIM proved to be fairly efficient in modelling the mass-metallicity relation over the whole epoch considered (Romeo Velon\`a et al. 2013), its drawback must not reside in anomalous metallicity variations along the RS, but rather in the stellar overproduction in high-mass galaxies at $z\gsim 1.2$, which was analyzed in Romeo et al. (2008).   

Focussing on the RS scatter (middle panels), we note that the intrinsic scatter in the HCS $(U-V)$ colour exhibits a wide range at $z>0.8$ ($0.01<(U-V)<0.25$), compared to both models, which instead converge in pointing at a lack of significant evolution of the scatter in all colours and environments.
As a comparison, the SAM of Menci et al. (2008) predicted a much higher intrinsic scatter at around 0.15 in $(U-B)$ (not shown here); however, given their large error bars as shown in C2016, their predictions still remain very marginally consistent with the higher-$z$ points of Mei et al. (2009) derived in the same colour.
A caveat here is that the $(U-B)$ colours at $B=-21.4$ mag in Mei et al. (2009) were systematically redder ($\sim 0.1$ mag) than those estimated in the HCS,
because these authors estimated their absolute magnitudes at a common redshift z=0.02 (see discussion in C2016, Section 5.1).

It can be seen that the RS scatter estimated in the HCS using all the galaxies down to the 90\% magnitude completeness limit (solid triangles), exhibits a mildly increasing trend with redshift which was already pointed out in C2016. However, as stressed by the authors, all the values are still consistent within the errors. When adopting instead a common absolute magnitude limit for all the clusters in the sample (empty triangles), any hint to an increasing trend with redshift disappears.
The absolute magnitude limits in the common limit sample are all close to the 90\% magnitude completeness limit in the HCS clusters at $z>1.1$, while are brighter, by up to 1.5 mag, than the 90\% completeness limit in the clusters at $z<1.1$: here the common limit sample is characterized by higher values of the intrinsic scatter compared to those derived with all the galaxies down to the 90\% magnitude completeness limit, suggesting that the latter may be underestimated due to the large photometric errors at faint magnitudes (ACS F850LP $>$ 23.0 AB mag) (See Fig. 3 in C2016). 

In order to test the effect of the photometric error on the estimate of the intrinsic scatter we ran the fit to the RS for the cluster RCS2319 ($z=0.91$) using the magnitude limit of the common limit sample and assuming that all the galaxies had the same photometric errors, regardless of their luminosity. We found that in the case in which we assumed a common photometric error of 0.01 mag for all the galaxies, the intrinsic scatter in the observed frame was $\sigma_c = 0.151 \pm 0.004$, while assuming a common error of 0.06 mag, the intrinsic scatter was $\sigma_c = 0.030 \pm  0.007$ , significantly lower than the previous case.
The result of this test shows that the value of the intrinsic scatter derived with the method illustrated in Section 3, may be underestimated by a large photometric error. For this reason, in the rest of the paper, we will only adopt the RS scatter obtained in the common limit sample.

\begin{figure*}	
	\includegraphics[width=0.32\textwidth]{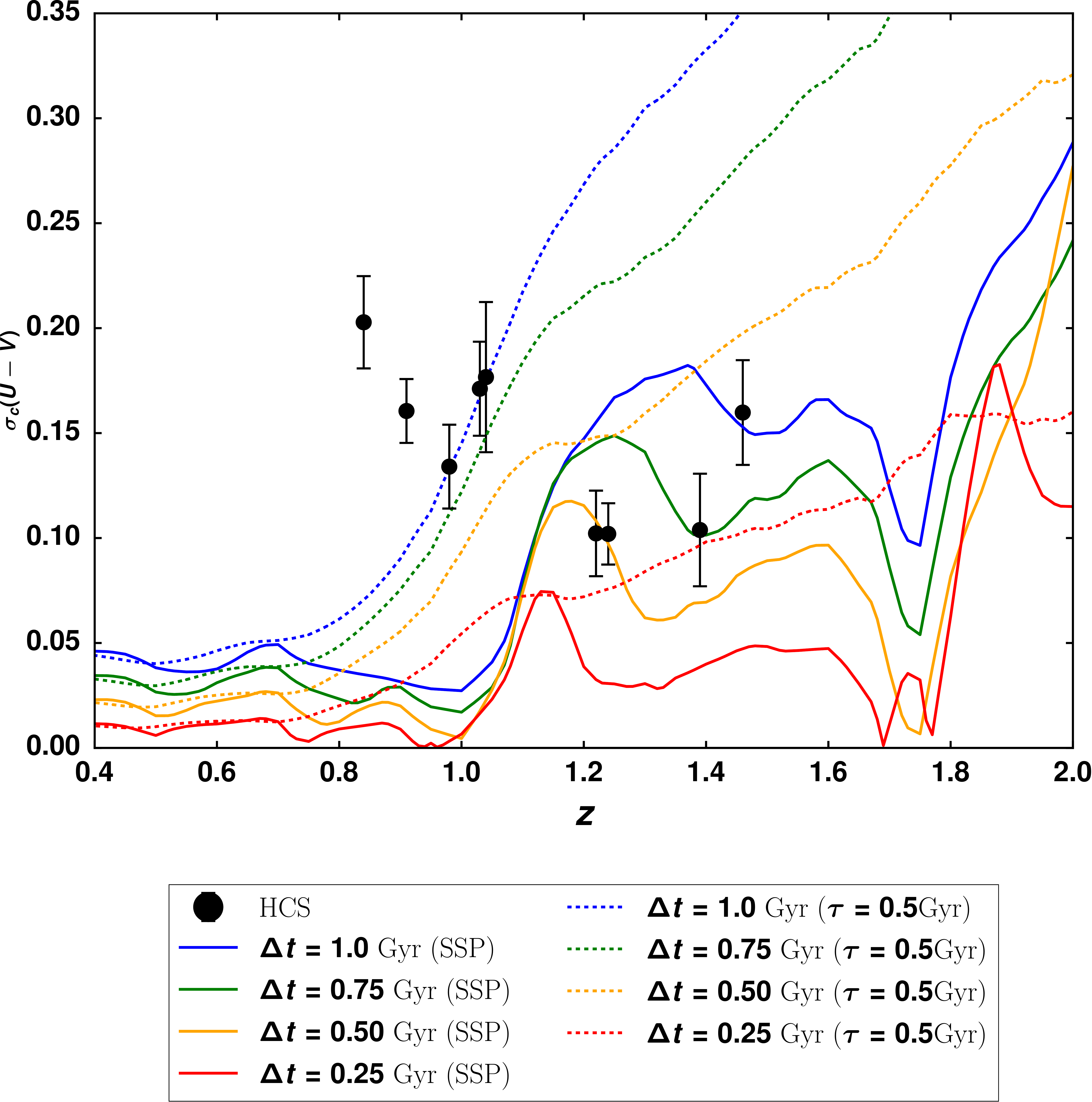}	
	\hfill
	\includegraphics[width=0.32\textwidth]{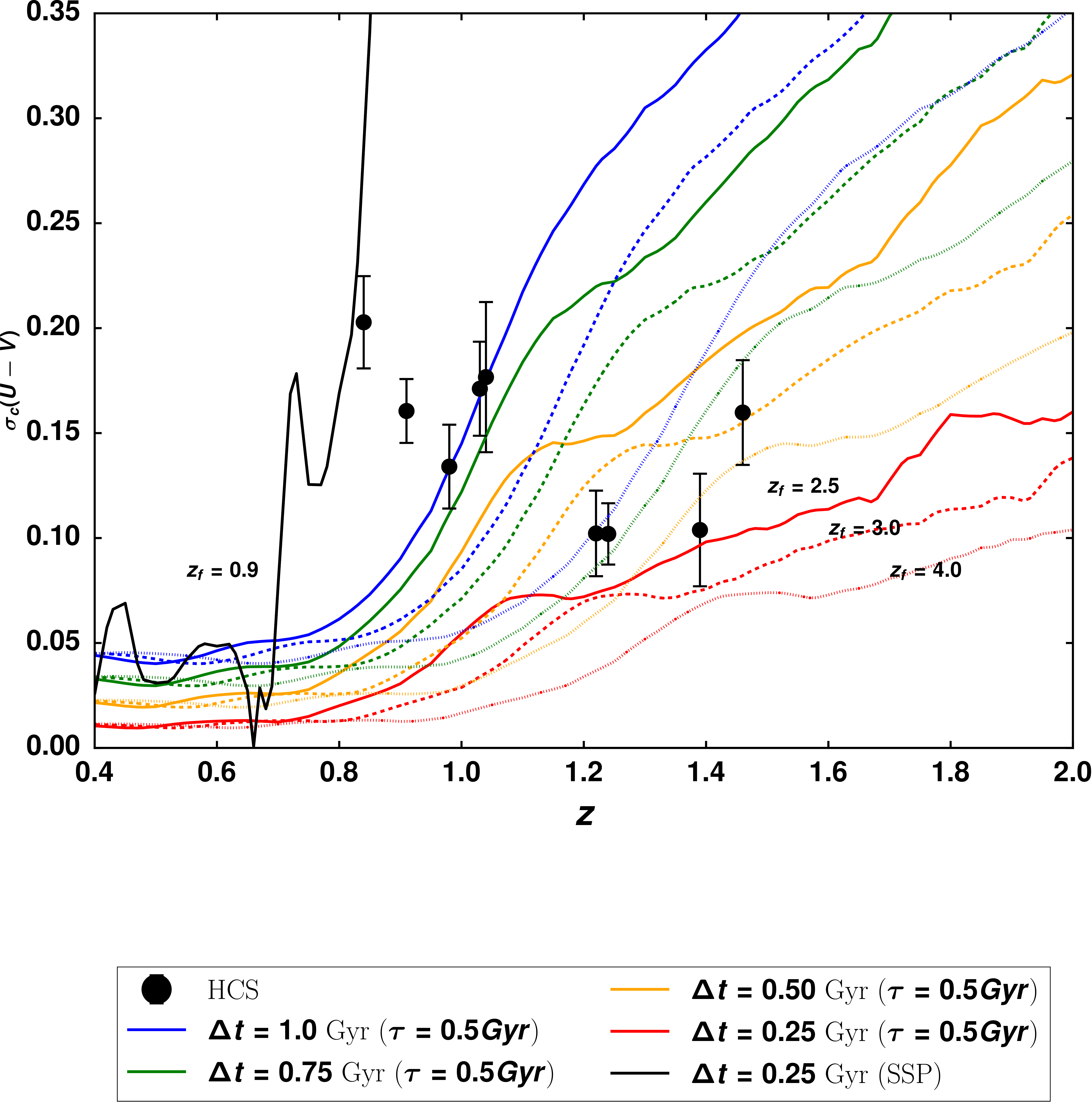}
	\hfill
	\includegraphics[width=0.32\textwidth]{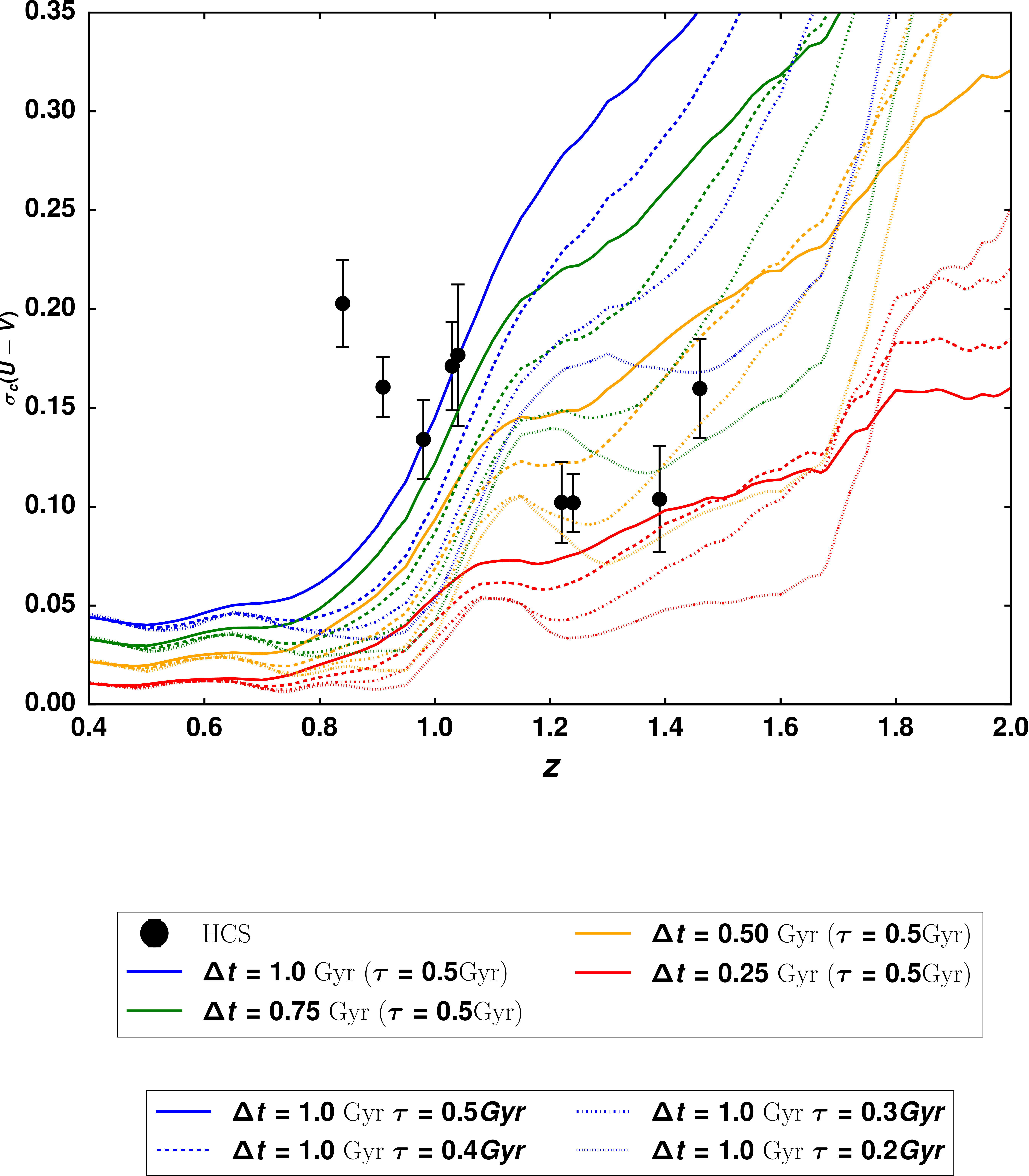}
        \caption{{\it a:} Dependence of the colour scatter on the SFH law; {\it b,c:} dependence of the colour scatter on the SFH $\tau$ and on the formation redshift of the youngest model, respectively, once fixed the SFH law.
Each curve represents the evolution of the $(U-V)$ colour scatter predicted for four age differences ($\Delta_t = 1, 0.75, 0.50, 0.25$ Gyr) and two SFH laws, namely instantaneous SSP (solid lines) and exponentially declining SFH (dotted lines), and is produced by calculating the difference between the rest-frame $(U-V)$ predicted by two models with identical SFH laws but formation ages differing by $\Delta_t$. With this method, the age difference $\Delta_t$ provides an estimate of the age scatter $\sigma_{\tau}$.}
\label{sigmac-z}
\end{figure*}

\subsection{Estimate of the Age Scatter}\label{sec:age_scatter_measurement_PC}

In this section we discuss the estimate of the age scatter of RS galaxies as inferred from the $U-V$ colour scatter $\sigma_c$. 
We only use here the $U-V$ rest-frame as this colour is more sensitive to stellar age effects than $U-B$ or $B-V$, which encompass regions that do not bracket the 4000 \AA\ break.
Several approaches have been used in the literature to infer the age range of RS galaxies from the colour scatter, and all of them were based on the comparison of the observer-frame or rest-frame $\sigma_c$ with the colours predicted by the passive evolution of stellar population models (e.g.: BLE92, Lidman et al. 2004, Papovich et al. 2010, Hilton et al. 2010, Foltz et al. 2015). The comparison of observed colours with the predictions of stellar population libraries relies upon the assumptions made on the model choice. Therefore, the adoption of only one of the methods that have been proposed in the literature may be affected by systematic effects resulting from such a choice. In order to control any systematic effect, we decided to adopt two different methods to infer the age scatter and compare the different estimates of $\sigma_{\tau} $obtained in each case.

We adopted solar metallicity stellar population synthesis models in our derivation of the age scatter, motivating our choice in Section 4.2.4, where we investigate the effects of metallicity variations in the RS shaping at redshift $z \sim 1$. However, we also account for the assumption on the model metallicity in the uncertainties on the average age $<\tau>$ (and formation redshift) and $\sigma_{\tau}$ obtained in Section 4.2.2 .

The models that we adopted for our derivations of $\sigma_{\tau}$ and $<\tau>$ of RS galaxies were either single stellar population (SSP) models or models with exponentially declining star formation histories (SFH). These two cases represent two extremes of the SFH of a stellar population, ranging from an extreme fast formation to a more gradual build-up of the stellar mass. Fig. \ref{sigmac-z}, which can be compared with Fig. 8 of Foltz et al. (2015), shows the evolutionary curves of the difference between the colors of stellar population models whose formation ages $\tau$ differ by a given $\Delta_t$. The formation redshifts of the models were chosen so that the youngest of the two models had its star formation ended by redshift $z=1.46$. The curves of Fig. \ref{sigmac-z} can be considered as evolutionary curves for the color scatter of a galaxy population under the assumption that the color scatter is exclusively a result of the age scatter of the galaxies (see Introduction).

The comparison between Fig. 2a and Fig. 2c shows that the effect of the e-folding time $\tau$ of the exponentially declining SFH on the predicted color scatter is that going from larger to smaller values of $\tau$ the shape of the evolutionary curve of the color scatter approximates that of the color scatter evolution of a SSP model.
In Fig. 2a we plot the evolutionary curves for the color difference of stellar population models with $\Delta_t = 1.0, 0.75, 0.50, 0.25$ Gyr. The exponentially declining SFH models were chosen with an e-folding time $\tau = 0.5$ Gyr and a minimum redshift $z=2.5$, which insured that the youngest model became passive by $z=1.46$ (the redshift of the most distant HCS cluster). It can be seen that at redshift $z\gsim 0.6$ the color scatter is highly dependent on the adopted SFH, with exponentially declining SFH models having larger color scatters than SSP models. We also note that similar values of the color scatter can be reproduced by either a SSP with a large $\Delta_t$ or an exponentially declining SFH with a small $\Delta_t$. This particularly affects the highest-redshift clusters in the HCS, at $z>1.2$, while for clusters at lower redshifts, the exponentially declining SFHs tend to better reproduce the observed color scatter.

We note, however, that the curves were built in such a way that the youngest model of each couple was passive by $z=1.46$, which is the redshift of the most distant cluster in the HCS. As it can be seen in Fig. 2b, the color scatter of the cluster RX0152 ($z=0.84$) ca also be reproduced by two SSP models with $\Delta_t =0.25$ Gyr and with the formation redshift of the youngest model $z_f = 0.9$.
Fig. \ref{sigmac-z} shows that, although the shape of the SFH law and the formation redshift of a stellar population have an effect at $z\gsim 0.6$, below this redshift the evolution of the color scatter appears flat and $\sigma_c = 0.05$ mag. However, as shown in Fig. 2b, the shape of the color scatter evolutionary curve is constrained by the age difference of the models and also by the formation redshift of the youngest model. Galaxies which quench their star formation at redshifts $z<0.8$ joining the RS will increase the age difference at the same time lowering the age of youngest galaxy. These effects both contribute in increasing the resulting color scattering.

\subsubsection{Method 1: direct estimate from observer-frame $\sigma_c$}\label{sec:age_scatter_method_1_PC}

The first measurement of the age scatter that we carry out follows the technique presented in Lidman et al. (2004) and consists in using the observer-frame RS to infer the age scatter of RS members. We take the observer-frame zero-point which, as mentioned above, can be considered as an average colour of a $L^*$ RS galaxy, and find the simple stellar population (SSP) model from the BC2003 library, with solar metallicity and Salpeter (1955) IMF which predicts the colour closest to that value. This measurement provides an estimate of the average age of RS galaxies. Then, by subtracting and adding to the zero-point the observed intrinsic scatter, we obtain an estimate of the colour of an average blue and red galaxy on the RS. We find also for these colours the BC2003 SSP models which best approximate them with the relative formation ages. These two values provide estimates of the ages of``typical'' young and old RS galaxies, respectively, while their difference provides an estimate of the age spread of RS galaxies. The age scatter $\sigma_{\tau}$ corresponds to half of this quantity.

\begin{figure}[t!]	
	\includegraphics[width=0.45\textwidth]{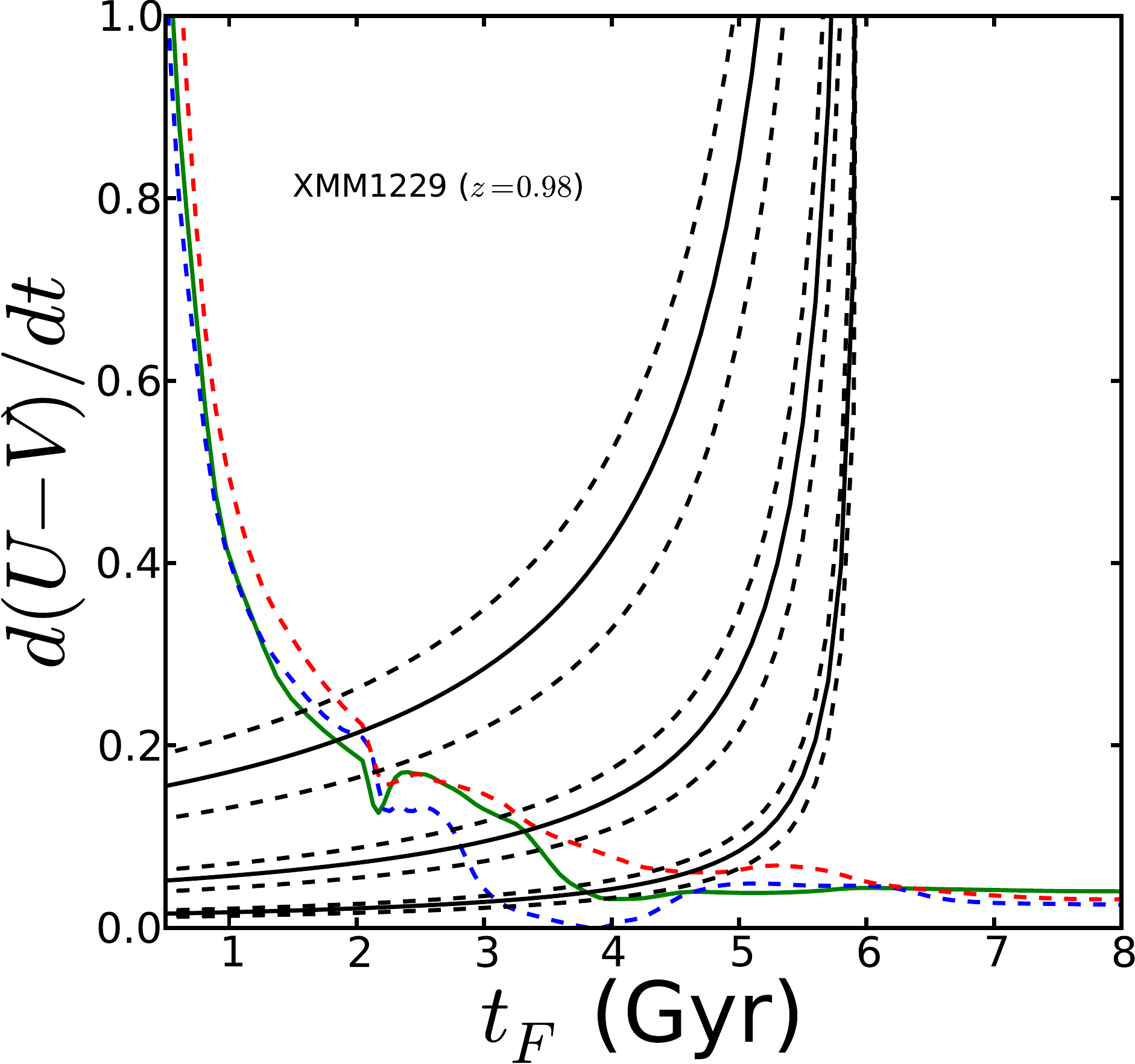}	
        \caption{The age-scatter test (third method) for the case of the cluster XMM1229 ($z=0.98$). The black solid curve represents the relation $\partial (U-V)_{t_F} / \partial t = \sigma_c / \left[\beta (t_H -t_F)\right]$, the black dashed lines represent the same relation at $\pm 1\sigma$ from the measured value of the intrinsic colour scatter. From left to right the curves represent the cases $\beta=0.1, 0.3, 1$. The green solid line is $\partial (U-V)_{t_F} / \partial t$ obtained from a single-burst BC2003 model with solar metallicity, while the blue and red dashed curves represent the same model but with metallicities $0.4 Z_\odot$ and $2.5 Z_\odot$, respectively. The intersection between the black curve and the model curve defines a lower limit to the lookback time to the last episode of star formation $t_F$. Here we assume $Z=Z_\odot$ and include the differences between the estimates of $t_F$ at $Z = Z_\odot$ and at $Z \neq Z_\odot$ in the uncertainty on this parameter.}
\label{dUVdt}
\end{figure}

\subsubsection{Method 2: The age-scatter test}\label{sec:age_scatter_method_3_PC}

The second method that we adopt to infer $\sigma_{\tau}$ is based on the emptirical relationship between the rest-frame color scatter, the age scatter, and the time of the end of star formation introduced in Bower, Lucey \& Ellis (1992, BLE92). 

If $t_F$ is the look-back time from the redshift of the cluster to the epoch at which galaxies ceased star formation and landed on the RS, $\Delta_t$ is the age spread of RS galaxies, and $U-V$ is the rest-frame $U-V$ color, then the rest-frame $U-V$ color scatter is related to these two quantities by the relation:
\begin{equation}\label{eq_1_PC}
\delta(U-V)_c = \frac{\partial (U-V)_{t_F}}{\partial t} \Delta_t
\end{equation}
where $\delta(U-V)_c$ is the $U-V$ color scatter of RS galaxies measured at the redshift of the cluster, and the derivative of the $U-V$ color with respect to cosmic epoch is evaluated at $t_F$. We can express the age scatter $\sigma_{\tau}$ as:
\begin{equation}\label{eq_2_PC}
\sigma_{\tau} = \frac{\Delta_t}{t_H - t_F} \times (t_H - t_F) = \beta (t_H -t_F)
\end{equation}
where $t_H$ is the cosmic age at the redshift of the cluster measured in Gyr.

BLE92 argued that the RS scatter can be used to constrain $\sigma_{\tau}$ because it poses an upper limit to the {\itshape{true}} color scatter of early-type galaxies. Thus one has:
\begin{equation}\label{eq_3_PC}
\delta(U-V)_c = \frac{\partial (U-V)_{t_F}}{\partial t} \beta (t_H -t_F) \leq \sigma_c.
\end{equation}
This relation allows one, given a value of $\beta$, to estimate $t_F$ and conversely, given a value of $t_F$, to estimate $\beta$ and hence $\sigma_{\tau}$. However, it does not allow one to estimate both quantities simultaneously. 

In order to break this degeneracy, following Jaff\'e et al. (2011, hereafter J11), we first estimated $t_F$ for several values of $\beta$ and then used the RS zero-point, which provides an estimate of the average RS color, to find the value of $\beta$ that best agrees with the observations. We estimated $t_F$ for $n \geq 10$ values of $\beta$ ranging from 0.1 to 1.0. The latter value corresponds to the case of totally asynchronous RS build-up, that is, galaxies on the RS formed across the entire $t_H - t_F$ range. The estimation of $t_F$ is illustrated in Fig. \ref{dUVdt} in the case of the cluster XMM1229 at $z=0.98$ and for the three cases $\beta =0.1, 0.3, 1.0$. We note that larger values of $\beta$ imply higher $t_F$ and, as a result, higher formation redshifts. Thus in the formation scenario of Eq. \ref{eq_3_PC} higher formation redshifts result in a more gradual build-up of the RS. In this work we explore the range $0.1 < \beta < 1.0$ to find the combination of $\beta$ and $t_F$ which best reproduces the observed RS scatter. 

\begin{figure*}	
\includegraphics[width=0.52\textwidth]{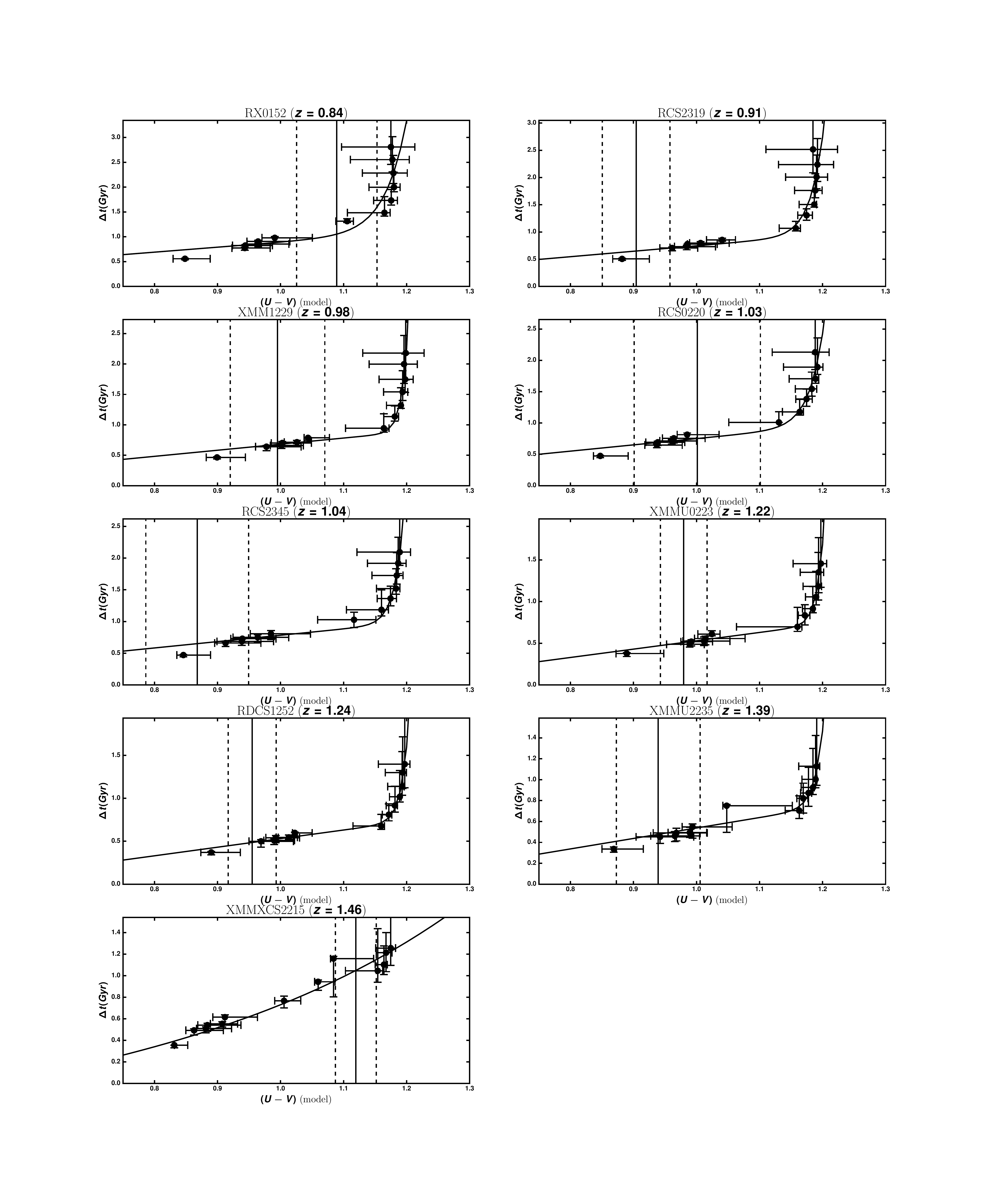}
\hfill
\includegraphics[width=0.52\textwidth]{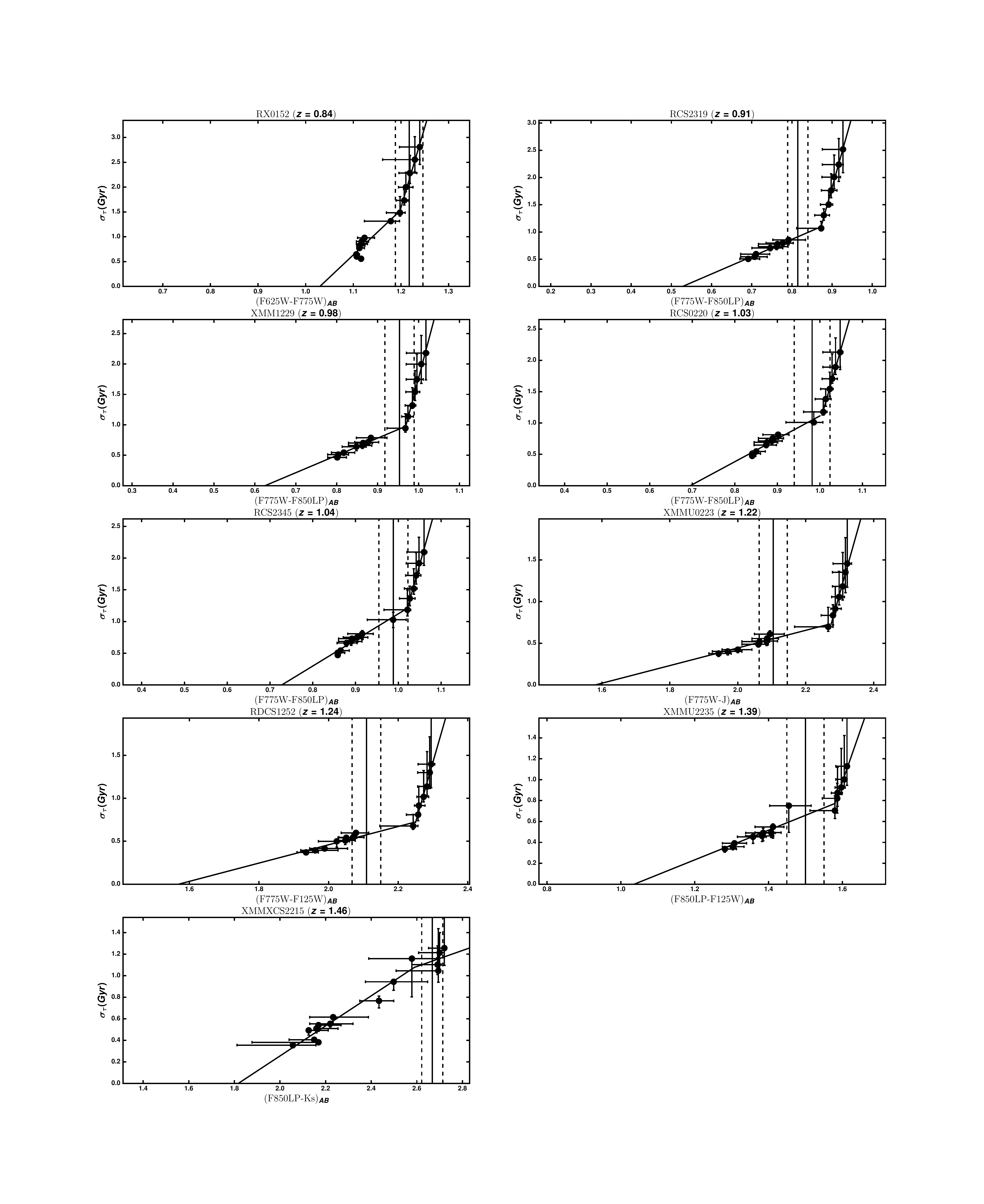}
        \caption{The $\Delta_t$ vs model colour relation in the HCS clusters: rest-frame $U-V$ colours obtained for models with exponentially declining SFH ({\it left}) and SSP ({\it right}) and formation redshift determined by $z_F$ and $\beta$ using Eq. \ref{eq_2_PC} as discussed in Section \ref{sec:age_scatter_method_3_PC}. The models were generated from the BC2003 library, adopting a Salpeter IMF and $Z_{\odot}$. The error bars on $\Delta_t$ and the $U-V$ colors take into account the effect of varying the metallicity with respect to the solar value. The vertical solid line coresponds to the rest-frame $U-V$ RS zero-point, estimated at $M_V^*$, while the dashed lines represent the zero-point plus or minus its uncertainty. Each point in the plot corresponds to a combination of $\Delta_t$ and $U-V$ colour obtained for $0.1 \leq \beta \leq 1.0$. The solid black curve represents the best fit curve from Eq. \ref{eq_4_PC}.}
\label{sigmat-UV}
\end{figure*}

To this end we built stellar population models from the Bruzual \& Charlot (2003, BC03) library with solar metallicity and Salpeter (1955) IMF. The formation redshift $z_{form}$ for each model was estimated from $z_F$ and $\beta$. We generated two sets of synthetic stellar populations, one composed of only SSP and the other of models built with exponentially declining SFHs. For the first family of models, we set the redshift of formation $z_{form}$ at the midpoint between $z_F$ and $z_{start}$, the latter being the redshift corresponding to $t_H - t_F -\Delta_t$. This choice of the formation redshift reflects the fact that the midpont of the age range of RS galaxies reflects a mean stellar age for the RS population in a scenario in which all RS galaxies had a rapid star formation history. For the second set of models we set the formation redshift $z_{form} = z_{start}$ and let the model form stars until $z_F$. The choice of $z_{form}$ for the exponentially declining SFH models reflects the fact that in the scenario in which all the galaxies on the RS had an exponentially declining SFH, the stars were formed less rapidly and the stellar populations had a more gradual build-up. For each model we considered the redshift evolution of the predicted $U-V$ color and compared with the value of the RS zero-point. This procedure is illustrated in Fig. \ref{sigmat-UV} for both the exponentially declining SFH (left) and the SSP (right) models.

In Fig. \ref{dUVdt} we show the effect that the uncertainty on the RS scatter and the assumption on the model metallicity have on the estimate of $t_F$ and $\sigma_{\tau}$. We account for both effects in the uncertainties on the derived values of $\Delta_t$, $t_F$, and on the model colors that are compared with the observed RS zero-points (see Fig. \ref{sigmat-UV}). 

We analyzed the relationship between the predicted $U-V$ colors and $\Delta_t$ and noted that for the exponentially declining SFH models we could fit the relation:
\begin{equation}\label{eq_4_PC}
\Delta_t = a (U-V)^b + (U-V) + c
\end{equation}
while for the SSP models a broken straight line provided a better fit to the points (Fig. \ref{sigmat-UV}).

The intersection between this curve (or the broken straight line) and the HCS $(U-V)$ zero-point provides an estimate of $\sigma_{\tau}$. For both the exponentially declining SFH and SSP models we repeated the same exercise with the RS zero-points measured in the observer frame. This allowed us to see whether the conversion to rest-frame colors could introduce systematic effects in the determination of $\sigma_{\tau}$.

We note that in the two cases of the clusters RX0152 and XMM1229 analyzed with exponentially declining SFHs and observer-frame colors we were not able to fit any analytic function to the $\Delta_t$ vs model color relation. For the first of the two clusters no stellar population model was able to reproduce the observed RS zero-point as the predicted colors were systematically bluer; for XMM1229 we interpolated between the points closest to the observed RS zero-point. In the cases in which the observed RS zero-point was bluer than the color predicted for the $\beta = 0.1$ models, we set $\sigma_{\tau}$ to the values of $\Delta_t$ corresponding to $\beta=0.1$ and regarded it as lower limits. These values will be represented as symbols without error bars in Figs. 6, 7, and 8. 

\begin{figure*}	
	\includegraphics[width=0.55\textwidth,clip]{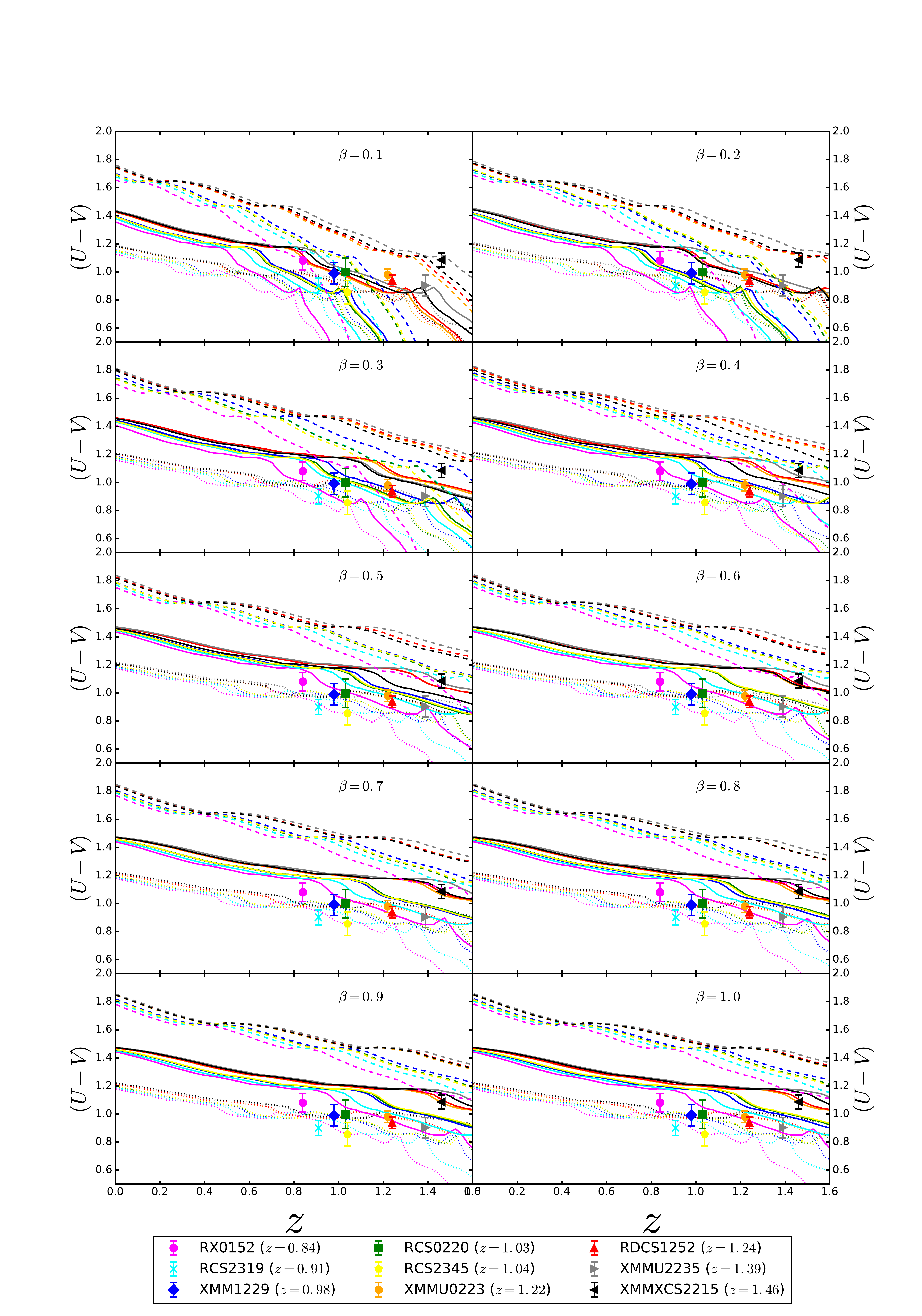}
	\hfill
	\includegraphics[width=0.55\textwidth,clip]{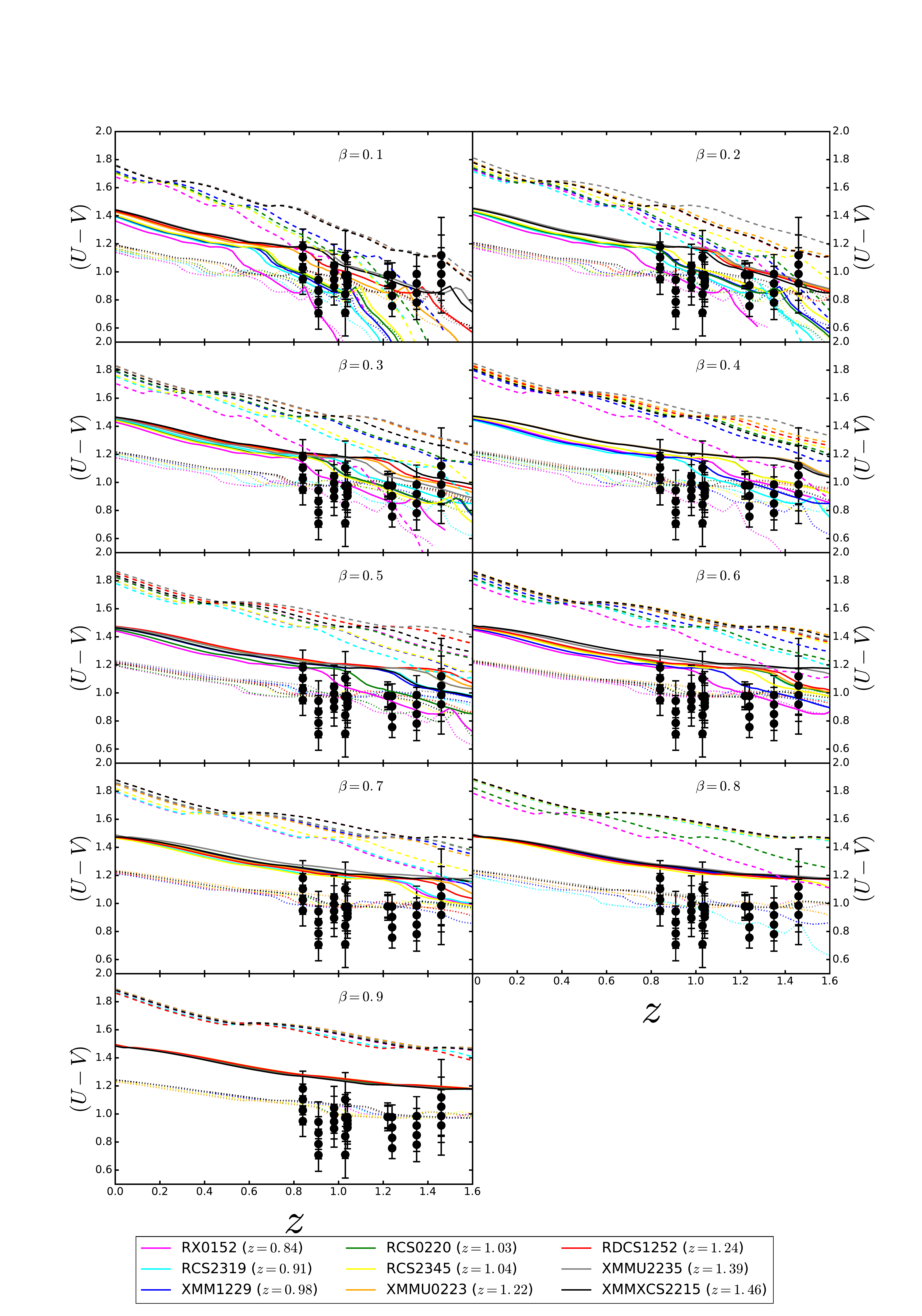}
        \caption{{\it a)}: Evolution of RS {\it U-V} zero-points predicted by BLE92 model: each panel refers to a value of the $\beta$ parameter, and each coloured curve in it is associated to a specific HCS cluster; solid curves correspond to BC03 models with $Z_{\odot}$, dotted with 0.4$Z_{\odot}$, dashed with 2.5$Z_{\odot}$. Points give intercept position of the RS at $M_V*$; {\it b)}: here points are the colors calculated on the RS using the slope and zero-points in Table 5 of C2016, at magnitudes $M_V = (-22.5,-21.5,-20.5,-19.5)$.}
\label{zerop}
\end{figure*}

\begin{table*}
\small
\addtolength{\tabcolsep}{-4pt}
  \caption{Summary of the measurements of the age scatter $\sigma_{\tau}$ and colour scatter $\sigma_c$ in HCS,
as in Fig. 6. From top to bottom $\sigma_{\tau}$ was estimated from direct comparison with the oberver-frame RS scatter (Section \ref{sec:age_scatter_method_1_PC}), BLE92 test with rest-frame $U-V$ RS zero-points and exponentially declining SFHs, BLE92 test with observer-frame RS zero-points and exponentially declining SFHs, BLE92 test with rest-frame  $U-V$ RS zero-points and SSP models, BLE92 test with observer-frame RS zero-points and SSP models (Section \ref{sec:age_scatter_method_3_PC}). The estimate of $\sigma_{\tau}$ for the cluster RX0152 with the BLE92 method, observer-frame RS zero-points and exponentially declining SFH did not converge and is reported as 0. $\sigma_\tau$ measured with the same configuration for the cluster XMM1229 was obtained by interpolating the $\Delta_t$ vs model color relationship and is highlighted by an asterisk.}
  \begin{minipage}{14 cm}
  \begin{tabular}{|c|c|c|}
    \hline
     \multicolumn{1}{|c|}{z} &  \multicolumn{1}{c|}{$\sigma_c \pm \delta \sigma_c$} &  \multicolumn{1}{c|}{$\sigma_{\tau} \pm \delta \sigma_{\tau}$} \\
     \multicolumn{1}{|c|}{}  & \multicolumn{1}{c|}{(mag)} & \multicolumn{1}{c|}{(Gyr)} \\                  
    \hline
     \hline
0.84 &    $0.20 \pm 0.02$        &     $4.0_{-0.6}^{+0.8}$         \\
     &                           &     $1.1_{-0.13}^{+0.5}$     \\
     &                           &     $0.0$      \\
     &                           &     $1.19_{-0.18}^{+0.18}$     \\
     &                           &     $2.1_{-0.7}^{+0.7}$     \\
                          
0.91 &    $0.161 \pm 0.015$      &     $2.9_{-0.04}^{+1.1}$            \\
     &                           &     $^*0.65_{-0.02}^{+0.05}$      \\
     &                           &     $0.74_{-0.013}^{+0.06}$      \\
     &                           &     $0.58_{-0.07}^{+0.11}$      \\
     &                           &     $0.91_{-0.08}^{+0.08}$     \\
   
0.98 &    $0.13 \pm 0.02$        &     $2.6_{-0.13}^{+0.8}$           \\
     &                           &     $0.68_{-0.08}^{+0.08}$      \\
     &                           &     $0.9_{-0.08}^{+1.3}$       \\
     &                           &     $0.68_{-0.14}^{+0.14}$     \\
     &                           &     $0.9_{-0.1}^{+0.6}$     \\
\hline
\end{tabular}
\hfill
  \begin{tabular}{|c|c|c|}
\hline
    \multicolumn{1}{|c|}{z} &  \multicolumn{1}{c|}{$\sigma_c \pm \delta \sigma_c$} &  \multicolumn{1}{c|}{$\sigma_{\tau} \pm \delta \sigma_{\tau}$} \\
     \multicolumn{1}{|c|}{} & \multicolumn{1}{c|}{(mag)} & \multicolumn{1}{c|}{(Gyr)} \\                  
    \hline   
1.03 &    $0.17 \pm 0.02$        &     $2.9_{-0.12}^{+1.0}$          \\
     &                           &     $0.75_{-0.10}^{+0.12}$     \\
     &                           &     $0.9_{-0.2}^{+1.0}$      \\
     &                           &     $0.79_{-0.18}^{+0.18}$     \\
     &                           &     $1.0_{-0.15}^{+0.5}$     \\
   
1.04 &    $0.18 \pm 0.04$        &     $3.0_{-0.15}^{+1.0}$          \\
     &                           &     $0.65_{-0.02}^{+0.08}$   \\
     &                           &     $0.8_{-0.11}^{+0.7}$     \\
     &                           &     $0.55_{-0.03}^{+0.16}$     \\
     &                           &     $1.07_{-0.14}^{+0.17}$     \\

1.22 &    $0.10 \pm 0.02$        &     $1.2_{-0.2}^{+0.5}$           \\
     &                           &     $0.51_{-0.04}^{+0.04}$     \\
     &                           &     $0.519_{-0.00017}^{+0.001}$     \\
     &                           &     $0.50_{-0.04}^{+0.04}$     \\
     &                           &     $0.55_{-0.04}^{+0.04}$     \\
\hline
\end{tabular}
\label{table2_PC}
\hfill
  \begin{tabular}{|c|c|c|}
    \hline
     \multicolumn{1}{|c|}{z} &  \multicolumn{1}{c|}{$\sigma_c \pm \delta \sigma_c$} &  \multicolumn{1}{c|}{$\sigma_{\tau} \pm \delta \sigma_{\tau}$} \\
     \multicolumn{1}{|c|}{}  & \multicolumn{1}{c|}{(mag)} & \multicolumn{1}{c|}{(Gyr)} \\                  
    \hline
     \hline

1.24 &    $0.102 \pm 0.015$      &     $1.3_{-0.12}^{+0.5}$          \\
     &                           &     $0.49_{-0.04}^{+0.04}$     \\
     &                           &     $0.515_{-0.0007}^{+0.004}$     \\
     &                           &     $0.48_{-0.04}^{+0.04}$     \\
     &                           &     $0.57_{-0.04}^{+0.04}$     \\

1.39 &    $0.10 \pm 0.03$        &     $1.4_{-0.1}^{+0.5}$            \\
     &                           &     $0.48_{-0.07}^{+0.07}$   \\
     &                           &     $0.69_{-0.11}^{+0.12}$    \\
     &                           &     $0.5_{-0.1}^{+0.1}$     \\
     &                           &     $0.66_{-0.07}^{+0.07}$     \\

1.46 &    $0.16 \pm 0.02$        &     $1.2_{-0.3}^{0.4}$          \\
     &                           &     $1.05_{-0.09}^{+0.10}$     \\
     &                           &     $1.16_{-0.07}^{+0.03}$     \\
     &                           &     $1.13_{-0.09}^{+0.03}$    \\
     &                           &     $1.14_{-0.03}^{+0.03}$     \\

     \hline
  \end{tabular}
\end{minipage}
\end{table*}

\subsubsection{Role of metallicity variations along the RS}

In principle the choice of metallicity should depend on the position on the RS. In our methods 1 and 3 we use as 
reference the intercept at $M_V*$ and in both cases we look for the model best reproducing the colour at that magnitude,
where the stellar mass of HCS clusters varies in the interval 10.7-11.0 dex (in solar log units).
Many works on the galaxy ZM relation report that within this mass interval metallicity oscillates between slightly sub-solar and slightly super-solar values around the epoch $z=$0.8-1.5 (e.g. Ferreras et al. 1999, S\'anchez-Bl\'azquez et al. 2009, Romeo-Velon\`a et al. 2013, Joergensen \& Chiboukas 2013, Gallazzi et al. 2014). Since we estimated the epoch of last episode of star formation by taking as reference the RS colour at $M_V*$, we believe that this is the base for an appropriate choice of solar metallicity in the models with methods 1 and 3 above discussed.

The curves in Fig. \ref{zerop}a represent the redshift evolution of SSP $U-V$ model colors derived for each HCS cluster with the BLE92 method (Section 4.2.2). Each panel corresponds to the value of $\beta$ indicated in the plots. The RS zero-points, estimated at $M_V^*$ for each cluster (color-coded symbols), appear to prefer models with solar or sub-solar metallicities for $\beta < 0.5$.

\begin{figure}[t!]	
	\includegraphics[width=0.45\textwidth]{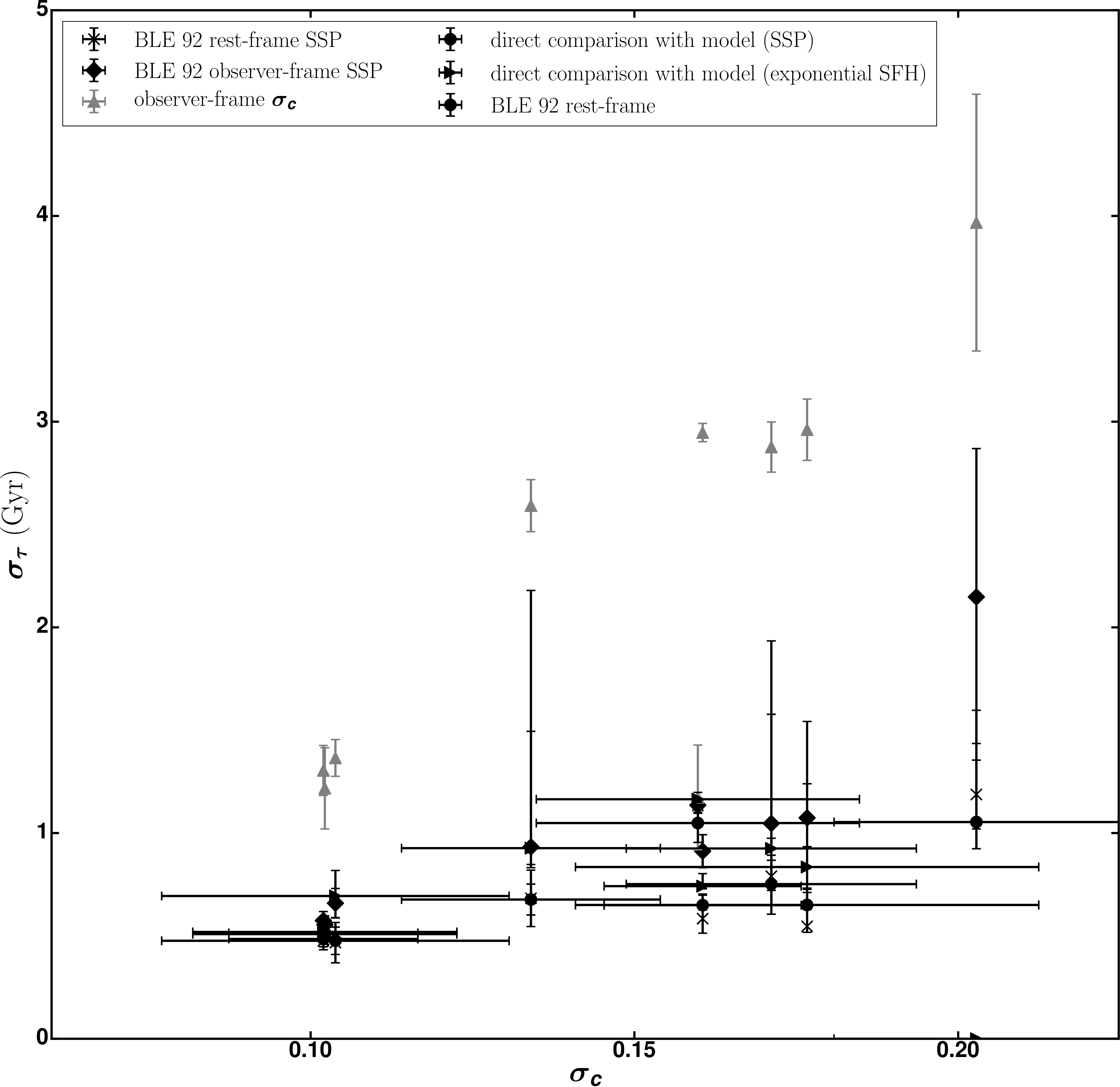}	
        \caption{Age scatter $\sigma_{\tau}$ measured in Gyr plotted as a function of rest-frame $U-V$ colour scatter.}
\label{sigmat-sigmac}
\end{figure}

Fig. \ref{zerop}b shows more clearly the relationship between galaxy color along the RS and stellar metallicity. The curves are the same as Fig. \ref{zerop}a; however the points with error bars represent the colors, calculated on the RS using the slope and zero-points in Table 5 of C2016, at magnitudes $M_V = (-22.5,-21.5,-20.5,-19.5)$. These values sample the range from the luminosity of the BCG down to the RS faint end in most HCS clusters. As expected, at bright magnitudes the RS colors are better approximated by higher-metallicity models. However, except for $\beta < 0.3$ and the cluster XMMXCS2215 ($z=1.46$), the RS colors still range between those predicted by solar and subsolar metallicities.

We can then conclude that our choice of adopting solar metallicity models in the inference of the mean age and age scatter of RS galaxies from the observed intrinsic RS scatter is well justified by the variation of colors observed along the RS. We find, indeed, that galaxy colors span the range predicted by models with solar or subsolar metallicities for a range of formation scenarios parametrized by the $\beta$ parameter introduced in Section 4.2.2. The RS color at $M_V^*$ is consistent with the predictions of solar metallicity models at $\beta > 0.3$.

\subsubsection{The $\sigma_{\tau}$ vs $\sigma_c$ Relation}\label{sec:age_scatter_discussion_PC}

Fig. \ref{sigmat-sigmac} shows the relationship between the age scatter $\sigma_{\tau}$ and the rest-frame $(U-V)$ color scatter, and Table 2 summarizes the results.

We find that the correlation between $\sigma_{\tau}$ and $\sigma_c$ changes according to the method adopted for the estimation of $\sigma_{\tau}$. In particular, we find that $\sigma_{\tau}$ appears well correlated with the RS color scatter (i.e. Spearman $\rho \geq 0.7$) when we infer $\sigma_{\tau}$ by both direct comparison with the observer-frame RS scatter (Method 1) and BLE92 test. The only case in which $\sigma_{\tau}$ and $\sigma_c$ are not correlated is when we infer the age scatter with the BLE92 method using the combination of exponentially declining SFH and observer-frame colors in the $\Delta_t$ vs model color relationship.

We note that the estimates of $\sigma_{\tau}$ obtained with Method 1 are systematically larger than those obtained with the BLE92 method. Interestingly, the estimates of the age scatter for the cluster RDCS1252 ($z=1.24$) obtained with this method are consistent with Lidman et al. (2004) and Rettura et al. (2011), who both obtained age spreads of RS galaxies greater than 1 Gyr for the same cluster. These estimates are also consistent with the average age spread inferred by Foltz et al. (2015) for RS galaxies in the SpARCS survey (Wilson et al. 2009, Muzzin et al. 2012). Fig. 6 shows that the estimates of $\sigma_{\tau}$ obtained with the BLE92 method and following different prescriptions on the stellar population models adopted for the $\Delta_t$ vs model color relationship are all consistent within the errors and lower than those obtained with Method 1, suggesting that either methods can be respectively biased toward higher or lower values of $\sigma_{\tau}$.

As discussed in the Introduction, the RS color scatter is the result of many competing physical factors of which the age spread and the mean age of the galaxies represent the major contributors (Tinsley 1972, Kodama \& Arimoto 1997, van Dokkum et al. 1998) and the metallicity spread contributes for up to 15\% (Nelan et al. 2005). Method 1 relies upon the assumption that the color scatter is only due to the age scatter (see discussion in Lidman et al. 2004), suggesting that this assumption may overestimate the actual contribution of the age spread to the observed RS scatter. We stress that the estimates of the age and age spread of RS galaxies derived from colors should be taken with caution and as only providing orders of magnitudes for these parameters. As illustrated in Worthey (1994), optical colors are degenerate stellar population indices, and the same variation in the observed color may be the result of a decrease in metallicity or an increase in stellar age (and vice versa). Nevertheless, the measurements of age and metal-sensitive spectral absorption indices in high-redshift ($z>0.8$) passive galaxies are challenging even with the largest 8-10m class telescopes as the uncertainties are not small enough to allow inferring the intrinsic age and metallicity scatters for RS galaxies (see e.g. Joergensen et al. 2014). Therefore, we remark the utility of using optical colors to infer the mean age and age spread of RS galaxies in distant clusters however warning that the preference of one method over the other to infer these parameters from the observed colors may lead to biased estimates. 

The conversion to absolute rest-frame colours is also a delicate task, and the assumptions made for the adopted set of stellar population models may lead to systematic differences among different works. Although C2016 discuss this in detail, we note that the consistency between the estimates of $\sigma_{\tau}$ obtained by using the observer-frame zero-points and those obtained by using the rest-frame ones with the same set of SSP models suggests that the conversion to rest-frame magnitudes is not responsible for the low $\sigma_{\tau}$.

\begin{figure}[t!]	
	\includegraphics[width=0.45\textwidth]{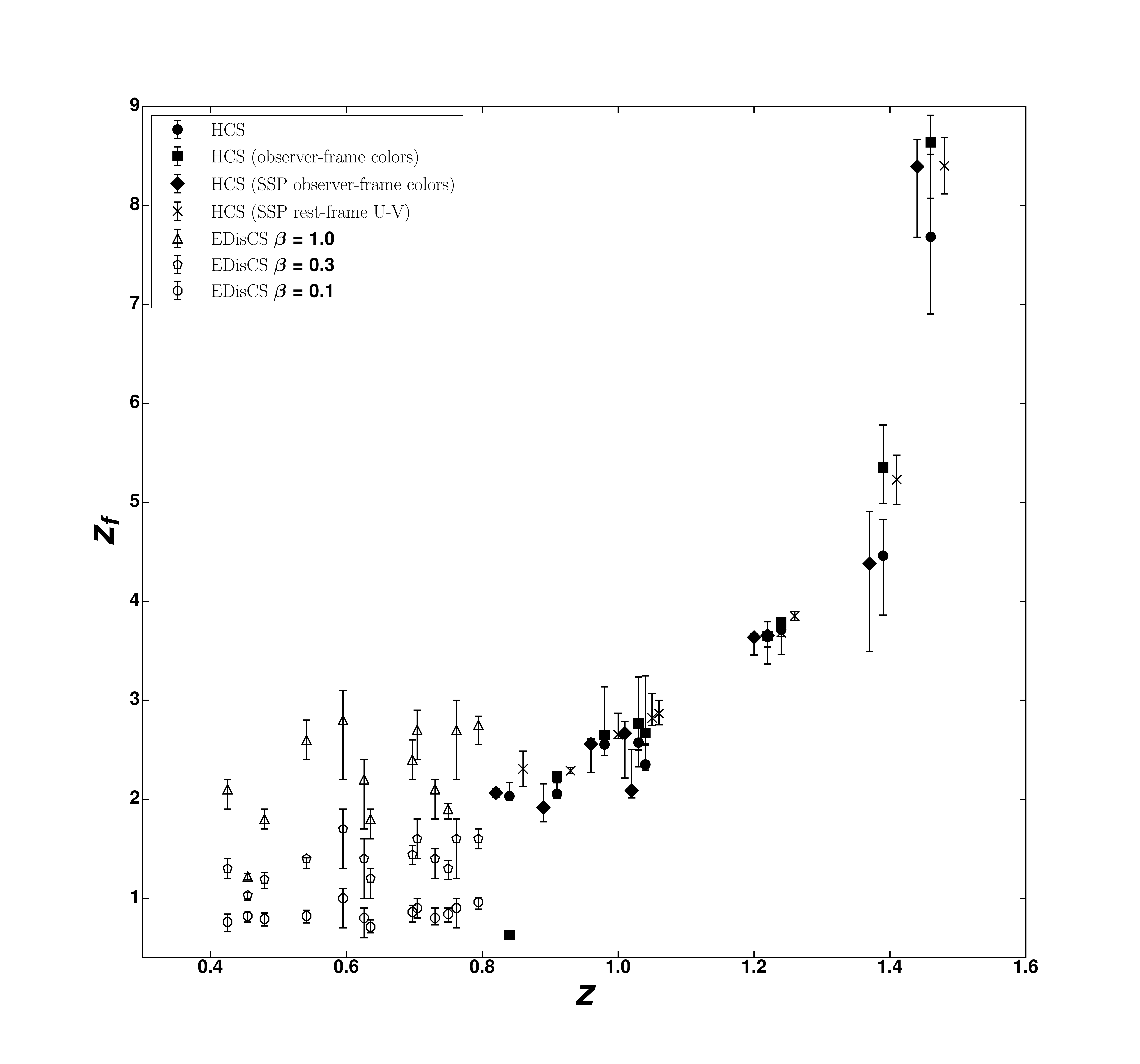}	
        \caption{Redshift of the end of RS formation $z_F$ plotted as a function of cluster redshift. The empty symbols correspond to the values reported in Table 2 of J11 for the three values $\beta=0.1, 0.3, 1.0$, while the filled symbols and the crosses correspond to the estimates obtained for the HCS clusters with the BLE92 test and the different model set-ups shown in the legend. $z_F$ increases with redshift, and the values found in HCS tend to be higher than those derived in J11 for the EDisCS sample.}
\label{zf-z}
\end{figure}

\section{Discussion}
\subsection{Analysis of RS building in the HCS}
In Fig. 2 we had noted that at redshifts $z>0.8$ the $(U-V)$ colour scatter is strongly dependent on the star formation history of galaxies. In particular, we note that at redshift $z=1.4$, which is about the redshift of the cluster XMMU2235, the same value of $\sigma_c$ can be yielded by two SSPs with starting 
$\sigma_{\tau} = 0.75$ Gyr and by two models with exponentially declining SFR with $\sigma_{\tau} = 0.25$ Gyr. Foltz et al. (2015) and C2016 also show that there is a large spread in $\sigma_c$ at $z>0.8$, probably reflecting the heterogeneity of clusters and proto-clusters at those epochs, which may be, in most cases, still assembling their Red Sequences.

On the opposite, at the redshifts of EDisCS, all the models represented in Fig. 2 predict $\sigma_c < 0.05$ Gyr. Interestingly, the colour scatter, measured in J11 in the observer-frame $(V-I)$ and $(R-I)$ Vega colours, approximately corresponding to rest-frame $(U-V)$ at $0.4<z<0.8$, are all larger than $\sigma_c = 0.05$, suggesting that these systems are younger than the HCS clusters. Indeed, large colour scatters would be expected in younger cluster, which are still assembling their Red Sequences. This can be observed, for example, in XMMXCS2215, which has the largest $\sigma_c$ in the HCS sample and has been shown to host actively star-forming galaxies in its core (Hilton et al. 2010). J11 detected a blue tail in the RS of EDisCS, mostly populated by S0 galaxies. Although the authors excluded these galaxies from their analysis, C2016 noted that the EDisCS clusters have halo masses lower than those that would be expected for the descendants of the HCS clusters at $0.4 < z < 0.8$ in a $\Lambda$CDM universe (see Fakhouri et al. 2010, Chiang et al. 2013,
Correa et al. 2015a, Correa et al. 2015b, Correa et al. 2015c)\footnote{Halo masses were measured in these clusters using the value of the velocity dispersion published in Halliday et al. (2004) and Milvang-Jensen (2008).}. This suggests that the EDisCS sample is composed of low-mass clusters with a recently assembled RS.

The values of $z_F$ for EDisCS and HCS support this argument. In Fig. \ref{zf-z} we plot $z_F$ as a function of cluster redshfit for the HCS and EDisCS, while Table \ref{table3_PC} summarizes the measurements of $z_F$ for all the model set-ups used in the BLE92 test. The EDisCS estimates are taken from Table 2 of J11, where they were derived for the three values of $\beta=$0.1 (empty circles), 0.3 (empty pentagons), and 1.0 (empty triangles). As expected from Eq. 2, the values of $z_F$ increase as a function of $\beta$: larger $\beta$ correspond to higher $z_F$. In agreement with what observed in J11, we find that $z_F$ increases with cluster redshift. However, we find a stronger increase with all model set-ups adopted in the present work compared to what can be observed in the EDisCS clusters for each value of $\beta$. The increase of $z_F$ with redshift was also discussed in J11, where the authors compared their estimates with those obtained from the values of $\sigma_c$ tabulated in Mei et al. (2009). They attributed this behaviour to the fact that the RS is acquiring new galaxies at lower redshifts and, therefore, the age at which its build-up was completed is also lower. This is an alternative view of the so-called {\itshape{progenitor bias}} (van Dokkum et al. 1998): as we observe clusters at higher redshifts, only the initial {\itshape{core}} of the lower-redshift RS is observable --which is composed of only the earliest galaxies which were formed in the cluster. 

J11 found that the scatter and zero-point of the EDisCS Red Sequences were best reproduced with $\beta \geq 0.3$. Except in the case $\beta = 1.0$, the values of $z_F$ for EDisCS are all smaller than those estimated in HCS. If all the EDisCS clusters were formed in the $\beta=1.0$ scenario, their $z_F$ would be higher than those estimated in the HCS clusters at $z<1.2$, but similar or lower than those estimated for the HCS clusters at $z>1.2$. However, we note that their Fig. 12 shows that, while the RS zero-points in clusters at $0.5 < z < 0.8$ are well reproduced by models with $\beta \geq 0.3$ (with the clusters at $z>0.6$ closer to the models with $\beta=1.0$), those at $z<0.5$ are closer to the colours predicted by the models with $\beta=0.1$. This suggests that most EDisCS clusters must have values of $z_F$ lower than those of the HCS clusters and, therefore, that their Red Sequences were assembled later. EDisCS is, therefore, composed of systems which are younger than those in HCS.

\begin{table*}
\small
\addtolength{\tabcolsep}{-5pt}
  \caption{Summary of the measurements of the redshift of end of RS formation $z_F$ using the BLE 92 method, as in Fig.7. From top to bottom in Col. 3 and 4 from the left are the estimates using models with exponentially declining SFHs and $U-V$ rest-frame RS zero-points, exponentially declining SFHs and observer-frame RS zero-points, SSP and $U-V$ rest-frame RS zero-points, SSP and observer-frame RS zero-points (see Section \ref{sec:age_scatter_method_3_PC} for details). $z_F$ measured with observer-frame RS zero-point and exponentially declining SFH in the cluster XMM1229 was obtained after interpolating the $\Delta_t$ vs model color relationship and is highlighted by an asterisk. $z_F$ for the cluster RX0152 in the same model/SFH configuration is reported as 0 because the estimate of $\sigma_\tau$ for this cluster did not converge.}
  \begin{minipage}{14 cm}
  \begin{tabular}{|c|c|c|c|}
    \hline
     \multicolumn{1}{|c|}{Cluster}  & \multicolumn{1}{c|}{z} &  \multicolumn{1}{c|}{$z_F$}  \\
    \hline
     RX0152        & 0.84 &    $2.03_{-0.05}^{+0.14}$ \rule{0pt}{2.6ex} \rule[-1.2ex]{0pt}{0pt}        \\
                   &      &    $0.0$       \rule{0pt}{2.6ex} \rule[-1.2ex]{0pt}{0pt}        \\
                   &      &    $2.07_{-0.04}^{+0.04}$     \rule{0pt}{2.6ex} \rule[-1.2ex]{0pt}{0pt}        \\
                   &      &    $2.31_{-0.18}^{+ 0.18}$     \rule{0pt}{2.6ex} \rule[-1.2ex]{0pt}{0pt}        \\

     RCS2319       & 0.91 &    $2.05_{-0.05}^{+0.11}$    \rule{0pt}{2.6ex} \rule[-1.2ex]{0pt}{0pt}        \\
                   &      &    $2.23_{-0.009}^{+0.02}$      \rule{0pt}{2.6ex} \rule[-1.2ex]{0pt}{0pt}         \\
                   &      &    $1.9_{-0.15}^{+0.2}$     \rule{0pt}{2.6ex} \rule[-1.2ex]{0pt}{0pt}        \\
                   &      &    $2.29_{-0.03}^{+0.03}$     \rule{0pt}{2.6ex} \rule[-1.2ex]{0pt}{0pt}        \\

     XMM1229       & 0.98 &    $2.55_{-0.11}^{+0.03}$    \rule{0pt}{2.6ex} \rule[-1.2ex]{0pt}{0pt}        \\
                   &      &    $^*2.7_{-0.03}^{+0.5}$       \rule{0pt}{2.6ex} \rule[-1.2ex]{0pt}{0pt}        \\
                   &      &    $2.6_{-0.3}^{+0.05}$     \rule{0pt}{2.6ex} \rule[-1.2ex]{0pt}{0pt}        \\
                   &      &    $2.7_{-0.04}^{+0.2}$     \rule{0pt}{2.6ex} \rule[-1.2ex]{0pt}{0pt}        \\
     \hline
  \end{tabular}
\hfill
  \begin{tabular}{|c|c|c|c|}
    \hline
     \multicolumn{1}{|c|}{Cluster}  & \multicolumn{1}{c|}{z} &  \multicolumn{1}{c|}{$z_F$}  \\
    \hline
     RCS0220       & 1.03 &    $2.6_{-0.2}^{+0.16}$       \rule{0pt}{2.6ex} \rule[-1.2ex]{0pt}{0pt}        \\
                   &      &    $2.8_{-0.3}^{+0.5}$       \rule{0pt}{2.6ex} \rule[-1.2ex]{0pt}{0pt}        \\
                   &      &    $2.7_{-0.5}^{+0.12}$       \rule{0pt}{2.6ex} \rule[-1.2ex]{0pt}{0pt}        \\
                   &      &    $2.8_{-0.07}^{+0.2}$     \rule{0pt}{2.6ex} \rule[-1.2ex]{0pt}{0pt}        \\

     RCS2345       & 1.04 &    $2.4_{-0.06}^{+0.2}$       \rule{0pt}{2.6ex} \rule[-1.2ex]{0pt}{0pt}         \\
                   &      &    $2.7_{-0.13}^{+0.6}$         \rule{0pt}{2.6ex} \rule[-1.2ex]{0pt}{0pt}       \\
                   &      &    $2.1_{-0.07}^{+0.4}$        \rule{0pt}{2.6ex} \rule[-1.2ex]{0pt}{0pt}         \\
                   &      &    $2.87_{-0.11}^{+0.13}$     \rule{0pt}{2.6ex} \rule[-1.2ex]{0pt}{0pt}        \\

     XMMU0223      & 1.22 &    $3.64_{-0.10}^{+0.04}$      \rule{0pt}{2.6ex} \rule[-1.2ex]{0pt}{0pt}          \\
                   &      &    $3.650_{-0.00016}^{+0.001}$        \rule{0pt}{2.6ex} \rule[-1.2ex]{0pt}{0pt}         \\     
                   &      &    $3.63_{-0.18}^{+0.04}$       \rule{0pt}{2.6ex} \rule[-1.2ex]{0pt}{0pt}          \\
                   &      &    $3.68_{-0.04}^{+0.04}$     \rule{0pt}{2.6ex} \rule[-1.2ex]{0pt}{0pt}        \\
     \hline
  \end{tabular}
\hfill
  \begin{tabular}{|c|c|c|c|}
    \hline
     \multicolumn{1}{|c|}{Cluster}  & \multicolumn{1}{c|}{z} &  \multicolumn{1}{c|}{$z_F$}  \\
    \hline
     RDCS1252      & 1.24 &    $3.7_{-0.3}^{+0.08}$         \rule{0pt}{2.6ex} \rule[-1.2ex]{0pt}{0pt}           \\
                   &      &    $3.787_{-0.0007}^{+ 0.004}$        \rule{0pt}{2.6ex} \rule[-1.2ex]{0pt}{0pt}          \\                           
                   &      &    $3.7_{-0.3}^{+0.14}$       \rule{0pt}{2.6ex} \rule[-1.2ex]{0pt}{0pt}           \\
                   &      &    $3.85_{-0.05}^{+0.05}$     \rule{0pt}{2.6ex} \rule[-1.2ex]{0pt}{0pt}        \\

     XMMU2235      & 1.39 &    $4.4_{-0.6}^{+0.4}$         \rule{0pt}{2.6ex} \rule[-1.2ex]{0pt}{0pt}           \\
                   &      &    $5.4_{-0.4}^{+0.4}$             \rule{0pt}{2.6ex} \rule[-1.2ex]{0pt}{0pt}           \\     
                   &      &    $4.4_{-0.9}^{+0.5}$        \rule{0pt}{2.6ex} \rule[-1.2ex]{0pt}{0pt}           \\
                   &      &    $5.2_{-0.2}^{+0.2}$     \rule{0pt}{2.6ex} \rule[-1.2ex]{0pt}{0pt}        \\

     XMMXCS2215    & 1.46 &    $7.7_{-0.8}^{+0.8}$            \rule{0pt}{2.6ex} \rule[-1.2ex]{0pt}{0pt}           \\
                   &      &    $8.6_{-0.6}^{+0.3}$        \rule{0pt}{2.6ex} \rule[-1.2ex]{0pt}{0pt}           \\
                   &      &    $8.4_{-0.7}^{+0.3}$         \rule{0pt}{2.6ex} \rule[-1.2ex]{0pt}{0pt}           \\
                   &      &    $8.4_{-0.3}^{+0.3}$     \rule{0pt}{2.6ex} \rule[-1.2ex]{0pt}{0pt}        \\

     \hline
  \end{tabular}
\end{minipage}
\label{table3_PC}
\end{table*}

\begin{figure}[t!]	
\includegraphics[width=0.45\textwidth,clip]{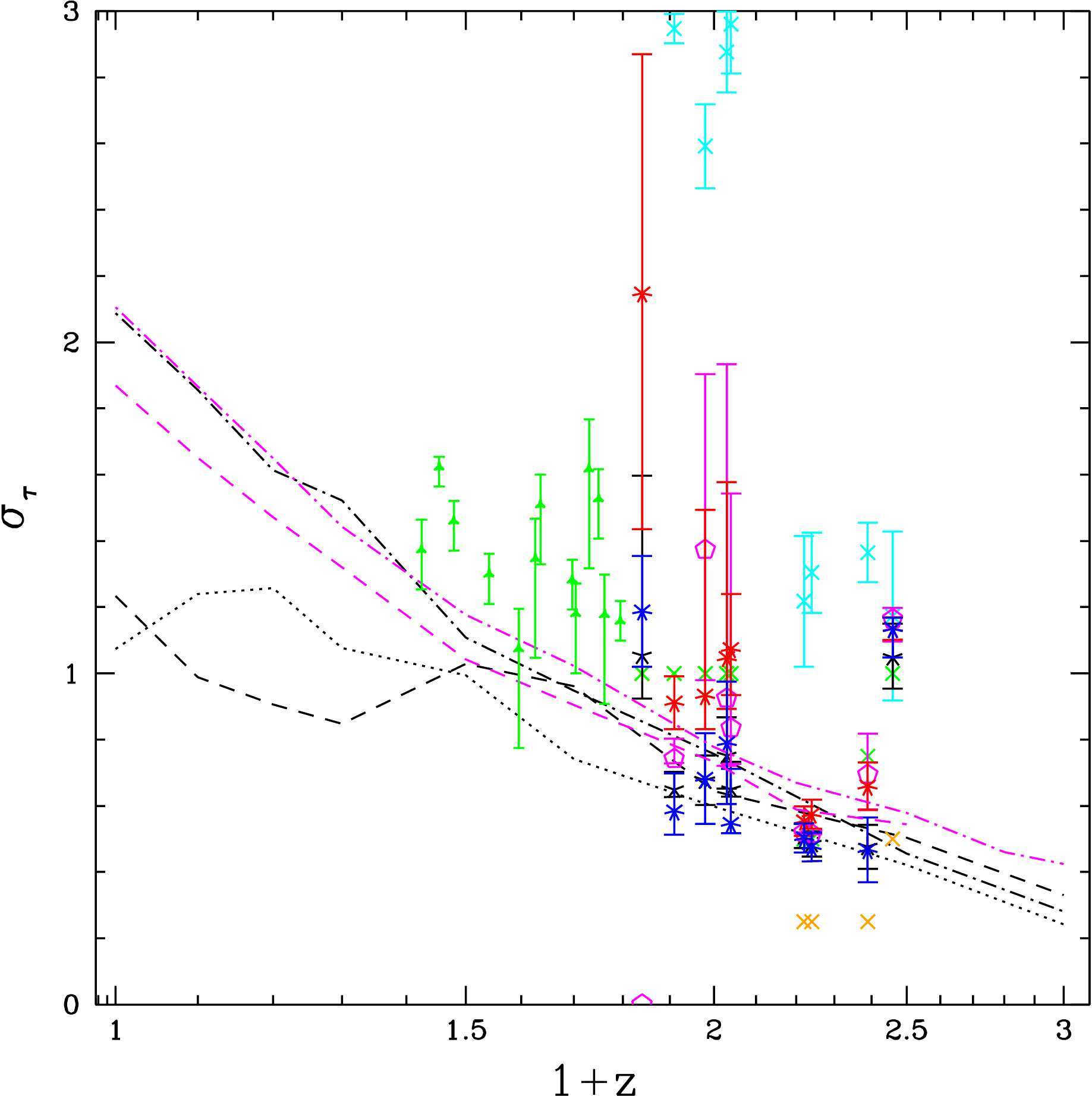}
\vfill
\includegraphics[width=0.45\textwidth,clip]{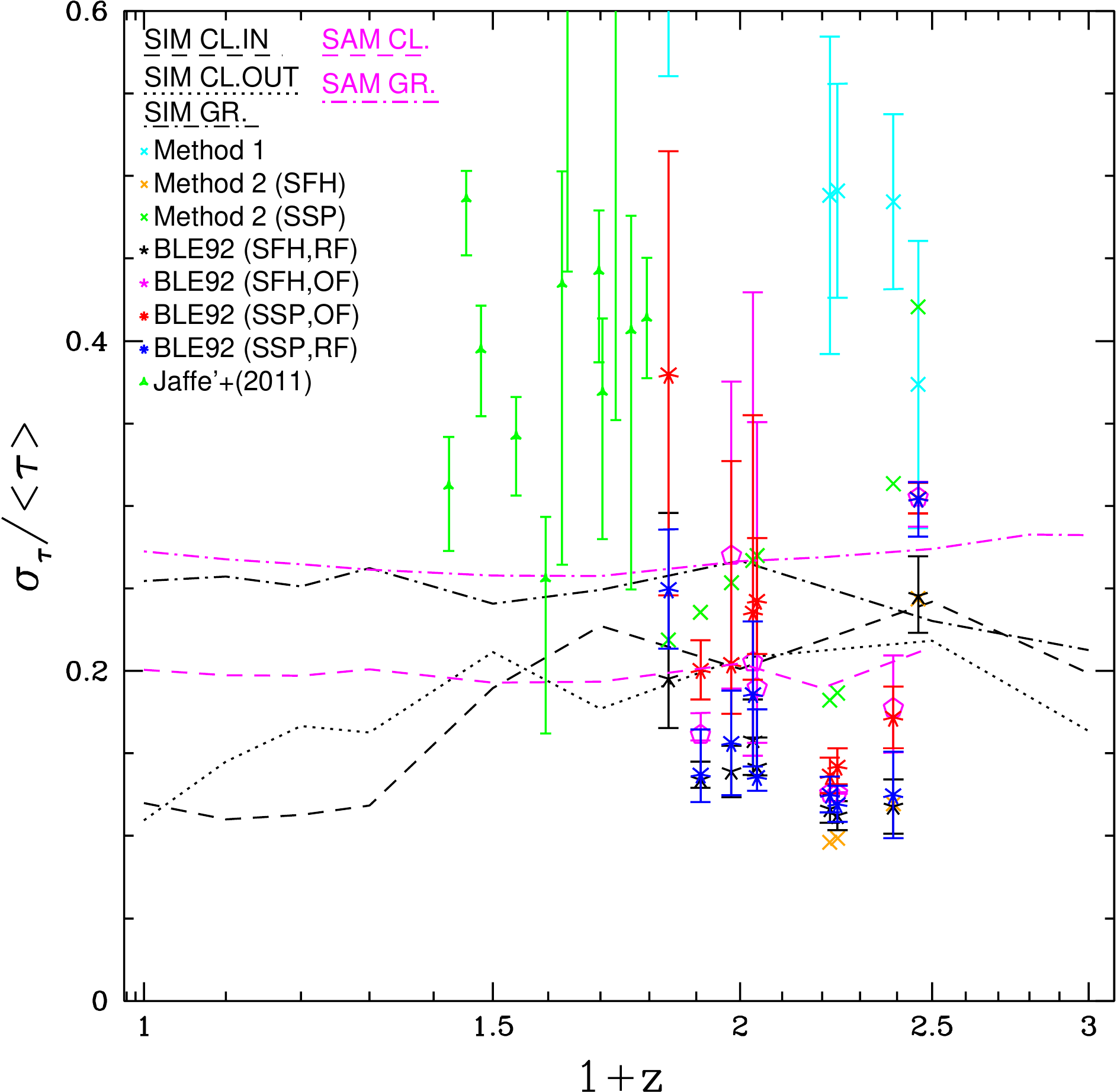}
\caption{Age scatter (above) and ratio of the age scatter over the mean age (bottom) as a function of redshift, from SIM and SAM (curves) and HCS (points), compared with data from J11. }
\label{scage}
\end{figure}

\subsection{Comparison with simulations}

As seen in van Dokkum et al. (1998) the evolution of RS colour scatter, that is of the colour differential of a
passively evolving stellar population, depends on the ratio $\sigma_\tau/<\tau>$, where $<\tau>$ is the average
stellar age weighted on luminosity of that population, and very little on the IMF and metallicity. In Nelan et al. (2005) and Gallazzi et al. (2006) the contribution of the metallicity scatter to the colour scatter is not negligible but lesser than the age scatter, at least in the local Universe. Hence we analyze here the ratio $\sigma_\tau/<\tau>$
as the main source of information for the RS global scatter.

In Fig. \ref{scage} the age scatter (upper) and the ratio $\sigma_\tau/<\tau>$ (lower) are plotted for our HCS, SIM and SAM samples, divided in clusters and groups, as a function of redshift. The trend for the former is steadily decreasing with $z$ for all SIM and SAM curves, except SIM clusters at low $z$. As for the $\sigma_\tau/<\tau>$ ratio, group galaxies follow a quasi-flat relation at considerably higher level both in SIM and SAM, while cluster galaxies (either in cores or outskirts)
are placed at roughly 0.1 dex lower. In general SAM curves show no redshift evolution of the ratio, while SIM clusters reach low values at $z\lsim0.5$. Except for this, it is interesting to note that SIM and SAM converge to a good match that overcomes the differences in the colour scatter seen in Fig. 1: this confirms that the larger spread in colours of the SAM galaxies was linked to their redder distribution and not to larger spread in ages with respect to SIM.
Indeed we tested that the RS $\sigma_\tau$ parameter alone is steadily increasing with time in both SIM and SAM (upper panel), due to the growing in number of the RS galaxies with different ages, and that only when divided by the average RS age we recover the quasi-constant trend as the colour scatter in Fig. 1. 

All in all then our models agree in predicting an increasing $\sigma_\tau$ with cosmic time: in order for the ratio
$\sigma_\tau/<\tau>$ to maintain approximately constant one needs to infer that $\sigma_\tau$ and $<\tau>$ be
proportional and hence $<\tau>$ increase with cosmic time too. On the other side, observational data suffer from 
the large uncertainty on all HCS measures as a consequence of the many assumptions we had to make upon the model parameters (metallicity, SFH, IMF, formation redshift, etc.). 
In general, the BLE92 test method yields lower values of $\sigma_\tau/<\tau>$, irrespectively of using either the {\it U-V} or the observer-frame zero-points (both with SSP), as well as using either SSP or exponential SFH in the rest-frame. Method 1 on the opposite yields higher values (as well as larger errorbars), likely because attributes the colour scatter to $\sigma_\tau$ alone, whilst the BLE92 method takes into account the whole ratio $\sigma_\tau/<\tau>$
(see Section 4.2.4).

However, as a general feature, our observational data concur in a $\sigma_\tau$ at $z\sim$1 smaller than at lower
$z$, along with a $<\tau>$ that is smaller as well (see Fig. 7), yielding then a ratio $\sigma_\tau/<\tau>$ approximately constant within the errors.
From the evolutionary point of view this means that the RS grows by swallowing young galaxies which have ceased forming stars, producing an age range that widens in time, also by effect of the discussed progenitor bias.

\section{Conclusions}
In this paper we employ previsions from SIM and SAM models on the trend and values of the RS colour scatter on galaxy
clusters and groups and compare them with observations of clusters at $z\sim$1-1.5, together with other clusters at lower redshift from literature. Both SIM and SAM shape an approximatively constant trend of the ratio $\sigma_\tau/<\tau>$, while the age scatter alone $\sigma_\tau$ increases with cosmic epoch. Both results are confirmed by observations of clusters at $z>$0.4 either in HCS and other works, suggesting that the RS gets incorporating new young galaxies after these shut down their star formation, giving rise to a widening in the range of stellar ages (see also Roediger et al. 2016).
Besides the general trend however, we found that the HCS values of both $\sigma_\tau/<\tau>$ and $\sigma_\tau$ strongly depend on the hypotheses made to infer them from the RS colour scatter, which produces quite large error bars. 

The main results of our work can be summarized as follows:
\begin{itemize}
\item{The redshift evolution of RS parameters of HCS sample is fairly reproduced by our SAM for the slope (where SIM yields too high values at high $z$)
and by our SIM for both scatter and zero-point (which are instead too high in the SAM), in all the colour considered.}
\item{The redshift of formation of the RS $z_F$ increases with cluster redshift, with higher redshift clusters having built-up the RS at earlier cosmic ages (Fig. 7). 
Our values are higher than those found at lower redshifts by J11 in the EDisCS survey for $\beta<1$. However, J11 found that only the RS zero-points of clusters at $0.6 < z < 0.8$ are better reproduced by models with $\beta = 1.0$, supporting the notion that $z_F$ increases with cluster redshift as well as the fact that the EDisCS clusters may be younger systems compared to our HCS clusters.}
\item{Our simulations and SAM fairly reproduce a quasi-constant redshift evolution of the ratio $\sigma_\tau/<\tau>$ (Fig.8), implying that the
mean galaxy age increases in time at the same rate as the age scatter over the whole interval considered.}
\item{Moreover, both SIM and SAM converge in predicting a clear gap in the sigma/tau ratio between groups and clusters, being around 0.2 dex higher in the formers at any epoch. But comparing with Fig. 1, it can be inferred that such difference only in the SAM is imputable to colour scatter, while in SIM must origin from a combination of the latter and the spread in age of RS members at each redshift. 
}
\end{itemize}

The BLE92 test applied to HCS yields lower $\sigma_\tau$ than those derived in the lower-redshift EDisCS clusters.
These clusters have, on average, lower halo masses compared to the HCS clusters, which may suggest more recent assembly histories for these systems. The higher $\sigma_\tau$ inferred from the results of J11 would thus be a result of the younger age of the EDisCS clusters. 
Nevertheless, we cannot exclude an underestimation of $\sigma_\tau$ in our analysis resulting from the particular assumptions made on the stellar population models.

However, from Fig. 2 we can see how $\sigma_c$ is effectively a function of the SFH, for $z\gsim$0.6, while at $z\lsim$0.6 both an
exponential SFH and the SSP produce similar colour scatters, converging towards values of $\sigma_c < 0.05$.
This means that measuring $\sigma_\tau$ at redshift higher than around 1 is not a simple task and strongly depends on the SFH; this is also evident from Fig. 8 of Foltz et al. (2015), where the values of $\sigma_c$ predicted by stellar population models and those measured in clusters at $z>1$ span a broad range of values.

All in all it is not straightforward to discriminate between a low age scatter due to wrong hypothesis on SFH, and a picture in which the HAWK-I clusters
can intrinsecally have low age scatter.
Again, comparing our Fig. 2 with Fig. 8 of Foltz et al. (2015), is a sound conclusion for most of the clusters that the colour scatter is consistent 
with an age scatter within the range 0.25$<\sigma_\tau<$0.50 (unless using SSP).

A possible explanation would then be that at such high redshifts the Red Sequences that we are seeing could be the cores or seeds of the Red Sequences of lower-redshift clusters. This conclusion is corroborated by the trends observed in the SIM and SAM and illustrated in the top panel of Fig. 8. However, as discussed in Section 4, stellar colors are highly degenerate stellar population indices, and more age-sensitive spectral absorption features, such as the H-Balmer lines (e.g.:\ $H \delta$ and $H \gamma$) and iron lines, need to be used in order to break the age-metallicity degeneracy at high redshifts. Nevertheless, these observations are extremely difficult for passive, early-type galaxies at $z>0.8$, with uncertainties that do not allow reliable estimates of the intrinsic age and metallicity scatters with even the highest signal-to-noise spectra taken with 8-10 m class telescopes. Stellar colors remain therefore a viable alternative to infer at least the order of magnitudes of the age spread of RS galaxies and its redshift evolution until the next generation of 20-30 m class telescopes will become operative in the next decade. 

Our next step is to include galaxy trees in the catalogues of our SAM, that can keep memory of halo assembly history; in this way we can select clusters and
groups at $z$=0 and trace back the main progenitors along the chain. Thus we will be able to quantify the RS scatter variations as function of redshift, when
clusters (and groups) accrete galaxies from the field, avoiding any ``progenitor bias".

\section*{Acknowledgments}
We thank C. Lidman for providing and discussing the HCS data.\\
This work is supported by the 973 program (n. 2015CB857000, 2013CB834900), the Foundation for Distinguished Young Scholars of Jiangsu Province (n. BK20140050), the ``Strategic Priority Research Program {\it The Emergence of Cosmological Structure}'' of the CAS (n. XDB09010000) and by China Postdoctoral Science Foundation through grant n. 2015M570488 (A.R.). A.R. and K.Xi are also supported by the NSFC, through grants n. 11550110183 and 11333008, respectively. E.C. acknowledges financial support by Chinese Academy of Sciences Presidents' International 
Fellowship Initiative, Grant n. 2015PM054.
P.C. acknowledges the support provided by FONDECYT postdoctoral research grant no 3160375 and by the Swinburne Chancellor Research Scholarship and AAO PhD Scholarship during 2015. Calculations were partially performed on the gSTAR national facility at Swinburne University of Technology: gSTAR is funded by Swinburne and the Australian Governments Education Investment Fund; we thank the Swinburne supercomputing and ITS teams for their support.

\end{document}